\documentclass[12pt]{article}
\usepackage{geometry}
\usepackage{color}
\usepackage{textcomp}
\usepackage{amsmath}
\usepackage{amssymb}
\usepackage{graphicx}
\usepackage{wasysym}
\usepackage{esint}

\makeatletter

\newcommand*\LyXThinSpace{\,\hspace{0pt}}

\newcommand{\lyxaddress}[1]{
	\par {\raggedright #1
	\vspace{1.4em}
	\noindent\par}
}

\usepackage[pagebackref=false,colorlinks=true,linkcolor=blue,citecolor=blue,urlcolor=blue]{hyperref}
\date{}

\makeatother

\begin{document}

\title{\textbf{Aharonov-Bohm Effect in Generalized Electrodynamics}}

\author{C. A. M. de Melo,$^{\ensuremath{1,}}$\thanks{Corresponding author: cassius.melo@unifal-mg.edu.br}$\,$
B. M. Perez,$^{\ensuremath{1}}$ J. C. Sumire Esquia,$^{\ensuremath{1,2}}$
R. R. Cuzinatto$^{\ensuremath{1}}$}
\maketitle

\lyxaddress{$^{\ensuremath{1}}$ Instituto de Ciência e Tecnologia, Universidade
Federal de Alfenas, Rodovia José Aurélio Vilela, 11999, CEP 37715-400,
Poços de Caldas, MG, Brazil}

\lyxaddress{$^{\ensuremath{2}}$ Instituto de Física Teórica, São Paulo State
University, P. O. Box 70532-2, CEP 01156-970, São Paulo, SP, Brazil}

\begin{abstract}
The Aharonov-Bohm (AB) effect is considered in the context of Generalized
Electrodynamics (GE) by Podolsky and Bopp. GE is the only extension
to Maxwell electrodynamics that is locally {\normalsize{}U(1)}-gauge
invariant, admits linear field equations and contains higher-order
derivatives of the vector potential. GE admits both massless and massive
modes for the photon. We recover the ordinary quantum phase shift
of the AB effect, derived in the context of Maxwell electrodynamics,
for the massless mode of the photon in GE. The massive mode induces
a correction factor to the AB phase shift depending on the photon
mass. We study both the magnetic AB effect and its electric counterpart.
In principle, accurate experimental observations of AB the phase shift
could be used to constrain GE photon mass. 
\end{abstract}

\tableofcontents{}

\section{Introduction\label{sec:Introduction}}

Maxwell equations of electromagnetism make room for the description
of the electric field $\mathbf{E}$ and the magnetic field $\mathbf{B}$
in terms of the electric potential $\varphi$ and the vector potential
$\mathbf{A}$ \cite{Felsager}. This is the case due to electrodynamics'
invariance under the U(1) group of transformations, or phase transformations.
This gauge freedom reduces the six degrees of freedom in the vectors
$\left\{ \mathbf{E},\mathbf{B}\right\} $ to the four degrees of freedom
in the pair $\left\{ \varphi,\mathbf{A}\right\} $ \cite{GriffithsEletro}.
It is thus said that electromagnetic interactions is the gauge theory
for the U(1) group \cite{Moriyasu1983}.

The equivalent description of electrodynamics in terms of the fields
$\left\{ \mathbf{E},\mathbf{B}\right\} $ or in terms of the gauge
potentials $\left\{ \varphi,\mathbf{A}\right\} $ raises the question
of what option is more fundamental. To put it another way: Would the
potential $\varphi$ and $\mathbf{A}$ be a simple mathematical artifact
to facilitate the computations in some applications of electromagnetism
but with no concrete physical reality? Previous to the advent of quantum
mechanics and gauge theory, the answer would most certainly be that
$\mathbf{E}$ and $\mathbf{B}$ are the fundamental entities of the
electromagnetic interaction \cite{GriffithsQuantum}. However, the
theme proved itself much more subtle since QED also presents gauge
invariance in a complemented way which includes vector states in the
physical Hilbert space \cite{MeloPimentelPompeia2006}. 

In fact, in 1959 Aharonov and Bohm proposed an experimental setup
that could assess the effect of the potentials $\left\{ \varphi,\mathbf{A}\right\} $
over charged particles in regions where the fields $\left\{ \mathbf{E},\mathbf{B}\right\} $
would be zero \cite{AharonovBohm1959}. The idea was that the wavefunction
associated to the charged particles would undergo a phase shift due
to interaction with the potentials $\left\{ \varphi,\mathbf{A}\right\} $.
This phase shift, if measured, would attribute fundamental physical
role to the gauge potential $\left\{ \varphi,\mathbf{A}\right\} $.\footnote{The role of potentials in the Aharonov-Bohm effect is also debated
in Ref. \cite{Vaidman2012}.} This interpretation fits well within the gauge theory framework,
which theoretically describes interaction through the minimal coupling
prescription introduced by Weyl \cite{Weyl1918}, used by Yang and
Mills \cite{Yang-Mills1954}, and justified deductively by Utiyama
\cite{Utiyama1956}. The minimal coupling prescription recommends
the mapping of ordinary differential operators such as $\partial_{\mu}$
onto the covariant derivative $\nabla_{\mu}=\partial_{\mu}+A_{\mu}$
\cite{O-Raio-Que-O-Parta1997,Acevedo2018}. Herein, notice the appearance
of the four-potential $A_{\mu}=\left\{ \varphi/c,\mathbf{A}\right\} $
(not of the fields $\left\{ \mathbf{E},\mathbf{B}\right\} $ embedded
in the field-strength $F_{\mu\nu}$).

As early as 1960, Chambers already claimed to have measured Aharonov-Bohm
(AB) effect \cite{Chambers1960}. This sparked debate in the community,
which resisted to fully embrace the result if any due to experimental
shortcomings. Arguably, Chambers' experiment presented accuracy issues
\cite{Peshkin1981}. However, these challenges were overcame over
the years and, by the time Tonomura and others published their findings
in 1983 \cite{Tonomura1983,Tonomura1986}, the AB effect was already
accepted beyond any criticism; moreover, it has since been measured
by different experimental techniques, e.g. \cite{Matteucci1985,Allman1993,Mukherjee2018,Overstreet2022}.
The dual of AB effect was proposed by Aharonov and Casher in 1984
\cite{AharonovCasher1984}. Ref. \cite{Batelaan2009} is an excellent
introductory review of Aharonov-Bohm and Aharonov-Casher effects.

The original AB effect was proposed in the context of non-relativistic
quantum mechanics. However, it has being generalized to the relativistic
realm of Dirac equation \cite{Lee2004}. The AB effect was also made
impact on the study of field theory on curved spacetime; in particular,
people worked out the AB effect due to the action of cosmic strings
and other topological defects, cf. e.g. Ref. \cite{Bakke2020}. Recent
literature include other applications too, e.g. Refs. \cite{Mansoori2016}.

One key ingredient for the original AB effect was the U(1) gauge symmetry
exhibited by the electromagnetic interaction. One might think that
the only possible consistent theoretical description of this interaction---one
that is both relativistic compatible and U(1) gauge invariant---is
Maxwell electrodynamics. This is indeed the case if one restricts
oneself to field equations containing up to second-order derivatives
of the gauge potentials $\left\{ \varphi,\mathbf{A}\right\} $ (first-order
derivatives of the field-strength). If the latter possibility is relaxed
then a new avenue opens up to describe electromagnetism. This was
the avenue followed by Podolsky, starting from 1942 \cite{Podolsky1942,Podolsky1944,Podolsky1948}---see
also the contribution by Bopp \cite{Bopp1940}. Podolsky introduced
a generalization to Maxwell field equations that includes fourth-order
derivatives of the gauge potentials while, at the same time, keeping
the linearity of the related differential equations. In terms in terms
of $\mathbf{B}$ and $\mathbf{E}$, these generalized field equation
read:
\begin{equation}
\begin{cases}
\left(1-a^{2}\square\right)\left(\nabla\cdot\mathbf{E}\right)=\frac{1}{\epsilon_{0}}\rho\\
\nabla\cdot\mathbf{B}=0\\
\nabla\times\mathbf{E}=-\frac{\partial\mathbf{B}}{\partial t}\\
\left(1-a^{2}\square\right)\left(\nabla\times\mathbf{B}-\mu_{0}\epsilon_{0}\frac{\partial\mathbf{E}}{\partial t}\right)=\mu_{0}\mathbf{J}
\end{cases}\label{eq:FieldEqsGE}
\end{equation}
where
\begin{equation}
\square\equiv-\frac{1}{c^{2}}\frac{\partial^{2}}{\partial t^{2}}+\nabla^{2}\,,\label{eq:DAlembertian}
\end{equation}
with $c=1/\sqrt{\mu_{0}\epsilon_{0}}$. Notice that Maxwell equations
are recovered from (\ref{eq:FieldEqsGE}) in the limiting case where
Podolsky parameter $a$ is negligibly small. 

Podoslsky's Generalized Electrodynamics (GE) has an additional atractive
feature: It is U(1) gauge invariant. Indeed, the paper \cite{Cuzinatto2007}
demonstrated that Podolsky electrodynamics is the only U(1) gauge
theory leading to field equation in terms of $\left\{ \varphi,\mathbf{A}\right\} $
that linear and of fourth-order. There sure exist other alternative
theories of electromagnetism besides GE, but they are non-linear \cite{Plebansky1968,MeloMedeirosPompeia2015}---e.g.
Euler-Heisenberg electrodynamics \cite{Heisenberg1936} and Born-Infeld
electromagnetism \cite{Born1934a,Born1934b}---, or violate U(1)
symmetry---e.g. de Broglie-Proca model \cite{Proca1936,deBroglie1922,deBroglie1926,deBroglie1940,Luo2005,Goldhaber2010}.
Podoslky original motivations for GE included the attempt to eliminate
divergences of the static potential of the point charge\footnote{Refs. \cite{Frenkel1996,Frenkel1999} are enticing works related to
this point.} and to help and cure divergences appearing in the attempt to build
the quantum version of electrodynamics. Latter on Podolsky electrodynamics
was abandoned in the face of the usual techniques of QED. However,
the interest in GE was renewed partially because of its gauge invariance
(which is broken in ordinary QED's upon imposition of the regularizing
cutoff techniques) but also because of its feature of accommodating
a massive mode for the photon. In fact, Podolsky parameter ${\color{blue}a=\hbar\left(cm_{\gamma}\right)^{-1}}$
brings a massive mode to the photon alongside the regular massless
mode typical of the regular Maxwellian description---as a quick check
of Eq. (\ref{eq:FieldEqsGE}) reveal. Constraining this mass is the
subject of a number of papers, e.g. \cite{Cuzinatto2011,Bonin2010,Bufalo2014}.
There are applications applenty of GE in QED \cite{Bufalo2011,Bufalo2012,Bufalo2013,BorgesBaronesMelo2019},
classical topics of eletromagnetism \cite{GalvaoPimentel,Accioly1998,Ortega2014,Bonin2016,Brandt2016,Lazar2019},
and even in gravitation---in conection to BH physics \cite{Cuzinatto2018,Frizo2023}---and
cosmology \cite{Cuzinatto2017}.

The U(1) gauge invariant character of GE enables us to study AB effect
in this context. This is the main goal of this paper. Besides the
theoretical interest of the study of Ab effect in GE, there is in
principle experimental application. The AB effect could serve as a
tool to probe GE in the low-energy regime. Specifically, AB effect
could be a means to constrain $a$ and add a lower bound to Podolsky
mass; this is a possibility we would like to assess here.

The paper rest of the paper is organized as follows. Section \ref{sec:Magnetic-AB-effect}
develops the magnetic AB effect in the context of GE. The electric
version of AB effect is covered in section \ref{sec:Electric-AB-effect}.
Finally, section \ref{sec:Final-Comments} contains our final comments,
discussion and perspectives. The Appendix contains an deeper discussion
of the role of U(1) gauge invariance for AB effect in GE.

\section{Magnetic AB effect\label{sec:Magnetic-AB-effect}}

We start off by calculating the magnetic field produced by a long
solenoid in the context of Generalized Electrodynamics. A beam of
quantum particles will be later considered travelling around the solenoid
and will be affected by Podolsky's version of the gauge potential.
The phase shift induced in the two branches of the splitted beam gives
the AB affect in GE. 

\subsection{Magnetic field of a long solenoid in GE\label{subsec:B-Field}}

According to Ref. \cite{Podolsky1948}, the GE version for Ampère-Maxwell
equation is 
\begin{equation}
\left(1-a^{2}\square\right)\left(\nabla\times\mathbf{B}-\mu_{0}\epsilon_{0}\frac{\partial\mathbf{E}}{\partial t}\right)=\mu_{0}\mathbf{J}\,,\label{eq:AmpMaxPod}
\end{equation}
where the coupling constant ${\color{blue}a=1/m_{\gamma}}$ \textcolor{blue}{(in
natural units)} has dimensions of length and is related to the photon
mass ${\color{blue}m_{\gamma}}$ in Podolsky electrodynamics. In the
stationary case, the fields $\mathbf{E}=\mathbf{E}(\mathbf{r})$ and
$\mathbf{B}=\mathbf{B}(\mathbf{r})$ do not depend on the time coordinate
$t$, and Eq. (\ref{eq:AmpMaxPod}) reduces to:
\begin{equation}
\left(1-a^{2}\nabla^{2}\right)\nabla\times\mathbf{B}=\mu_{0}\mathbf{J}\,,\label{eq:AmpPod}
\end{equation}
where $\mathbf{J}$ is the current density. This is equivalent to:
\begin{equation}
\nabla\times\mathbf{B}_{\text{eff}}=\mu_{0}\mathbf{J}\label{eq:AmpDiffBeff}
\end{equation}
under the definition
\begin{equation}
\mathbf{B}_{\text{eff}}=\mathbf{B}-a^{2}\nabla^{2}\mathbf{B}\label{eq:Beff}
\end{equation}
which adds the Podolsky contribution $\left(-a^{2}\nabla^{2}\mathbf{B}\right)$
to the regular Maxwellian term. The later is recovered in the limit
$a\rightarrow0$. Eq. (\ref{eq:AmpDiffBeff}) has exactly the same
form of the traditional Ampère law (under the mapping $\mathbf{B}\rightarrow\mathbf{B}_{\text{eff}}$).
Therefore, its integral form will also be the familiar one: 
\begin{equation}
\varoint_{C}\mathbf{B}_{\text{eff}}\cdot d\mathbf{l}=\mu_{0}I_{\text{enc}}\label{eq:AmpBeff}
\end{equation}
In Eq. (\ref{eq:AmpBeff}) we see the definition of the current $I_{\text{enc}}$
puncturing through the surface $S$ enclosed by the circuit $C$:
$I_{\text{enc}}=\int_{S}\mathbf{J}\cdot\mathbf{n}dS$. The solution
to this equation for various currents configurations is the subject
of study of electromagnetism courses. For instance, a long solenoid
of radius $R$, length $l\gg R$ and number of loops $N$ carrying
a steady current $I$ produces an internal uniform magnetic field
equal to $\mu_{0}nI\hat{\mathbf{z}}$ in the direction $\hat{\mathbf{z}}$
along its axis; $n=N/l$ is the density of turns \cite{GriffithsEletro}.
Outside the solenoid the field is zero. This conclusion follows directly
from Eq. (\ref{eq:AmpBeff}), i.e.:
\begin{equation}
\mathbf{B}_{\text{eff}}=\begin{cases}
\mu_{0}nI\hat{\mathbf{z}}\,, & r<R\quad\left(\text{inside the solenoid}\right)\\
0\,, & r>R\quad\left(\text{outside the solenoid}\right)
\end{cases}\label{eq:Beff_solenoid}
\end{equation}
Notice that the effective field $\mathbf{B}_{\text{eff}}$ inside
the solenoid in constant---it assumes the same value irrespective
of the distance $r$ from the solenoid's axis. The actual field $\mathbf{B}$
is calculated by substituting (\ref{eq:Beff_solenoid}) into the definition
(\ref{eq:Beff}) and, from this perspective, the effective magnectic
field $\mathbf{B}_{\text{eff}}$ works as the source of $\mathbf{B}$. 

In principle, the magnetic field would be a vector field of the type
$\mathbf{B}\left(r,\phi,z\right)=B_{r}\left(r,\phi,z\right)\hat{\mathbf{s}}+B_{\phi}\left(r,\phi,z\right)\hat{\mathbf{\phi}}+B_{z}\left(r,\phi,z\right)\hat{\mathbf{z}}$
in cylindrical coordinates. However, from symmetry arguments and the
right-hand rule:\footnote{The angle $\phi$ localizes points around the solenoid rotated from an arbitrary defined axial direction $\hat{\mathbf{r}}$. Physically, it is not reasonable that \textbf{$\mathbf{B}$} changes if one rotates around the cylinder's axis, which looks the same from all directions perpendicular to the axis; then, \textbf{$\mathbf{B}$ }is not a function of $\phi$. By a similar token, $\mathbf{B}$ should not depend on
the coordinate $z$, which selects a particular point along the direction
parallel to the axis of the solenoid: For an idealized infinitely
long solenoid, all point are equidistant from the origin, which itself
is at an arbitrary position along the axis. Conclusion: From symmetry
arguments, we know that $\mathbf{B}$ can not depend on the coordinates
$\phi$ and $z$: $\mathbf{B}=\mathbf{B}\left(r\right)$. Moreover,
$\mathbf{B}$ will not have components pointing in the radial and
angular directions: in accordance to the right-hand rule applied to
the circular current distribution, $\mathbf{B}$ should point in the
$\hat{\mathbf{z}}$-direction: $B_{r}\left(r\right)=B_{\phi}\left(r\right)=0$. (Recall the Maxwell's case, where the magnetic field for the long
solenoid is given by $\mu_{0}nI\hat{\mathbf{z}}$ on the grounds of
symmetry; the same symmetries that apply in Podolsky's case.)}
\begin{equation}
\mathbf{B}=\mathbf{B}\left(r\right)=B_{z}\left(r\right)\hat{\mathbf{z}}\qquad\left(\text{symmetry}\right)\label{eq:SymmetryOnB}
\end{equation}
This greatly simplifies the form of the Laplacian of $\mathbf{B}$
in cylindrical coordinates. In fact, by making use of the dimensionless
coordinate
\begin{equation}
\zeta=\frac{r}{a}\label{eq:zeta}
\end{equation}
the differential equation (\ref{eq:Beff}) takes on the from:
\begin{equation}
\frac{d^{2}B_{z}}{d\zeta^{2}}+\frac{1}{\zeta}\frac{dB_{z}}{d\zeta}-B_{z}=-B_{\text{eff}}\,,\label{eq:DiffEqBz(zeta)}
\end{equation}
where $B_{\text{eff}}=\mu_{0}nI$ for $\zeta<R/a$ and $B_{\text{eff}}=0$
for $\zeta>R/a$, in accordance with Eq. (\ref{eq:Beff_solenoid}).
Outside the solenoid, the right-hand side of Eq. (\ref{eq:DiffEqBz(zeta)})
is null, in which case this differential equation is recognized as
the (homogeneous) modified Bessel equation of order zero---see e.g.
Ref. \cite{Abramowitz}, Eq. 9.6.1 with $\nu=0$. Its solution is
in terms of the modified Bessel functions $I_{\nu}\left(\zeta\right)$
and $K_{\nu}\left(\zeta\right)$: 
\begin{equation}
B_{z}\left(\zeta\right)=c_{1}I_{0}\left(\zeta\right)+c_{2}K_{0}\left(\zeta\right)\qquad\left(\zeta>\frac{R}{a}\right)\,.\label{eq:SolDiffEqModBesselOut}
\end{equation}
This should be valid everywhere provided that $r>R$, including the
limit of large $r$, i.e. large $\zeta$. In this regime, we see that
\cite{Abramowitz}:
\begin{equation}
I_{0}\left(z\right)\sim\frac{e^{z}}{\sqrt{2\pi z}}\Rightarrow\lim_{z\rightarrow\infty}I_{0}\left(z\right)\rightarrow\infty\,,\qquad\text{and}\qquad K_{0}\left(z\right)\sim\sqrt{\frac{\pi}{2z}}e^{-z}\Rightarrow\lim_{z\rightarrow\infty}K_{0}\left(z\right)=0\,.\label{eq:I0_K0_largez}
\end{equation}
Therefore, we must discard the term scaling with $I_{0}\left(z\right)$
if we want to avoid that the field $B_{z}$ diverges far from the
solenoid, i.e. we are forced to choose: $c_{1}=0$. This is the first
boundary condition that we impose to our physical configuration, in
the face of which, Eq. (\ref{eq:SolDiffEqModBesselOut}) reduces to:
\begin{equation}
B_{z}\left(\zeta\right)=c_{2}K_{0}\left(\zeta\right)\qquad\left(\zeta>\frac{R}{a}\right)\,.\label{eq:Bz(c2)out}
\end{equation}
The constant $c_{2}$ is still to be determined. This can be done
by imposing continuity of $B_{z}$ at the solenoid's surface, where
$r=R$, i.e. $\zeta=R/a$. For this goal, we need to match (\ref{eq:Bz(c2)out})
with its counterpart $B_{z}\left(\zeta\right)$ valid inside the solenoid.
So we naturally turn to the task of solving the non-homogeneous Eq.
(\ref{eq:DiffEqBz(zeta)}). Since this is a second-order differential
equation, its general solution is the sum $B_{z}=B_{z}^{\text{hom}}+B_{z}^{\text{part}}$,
for $\zeta\leqslant R/a$. Here, $B_{z}^{\text{hom}}$ indicates the
solution to the homogeneous modified Bessel equation; thus: $B_{z}^{\text{hom}}\left(\zeta\right)=c_{3}I_{0}\left(\zeta\right)+c_{4}K_{0}\left(\zeta\right)$.
(We put in other constants, $c_{3}$ and $c_{4}$, for generality.)
The term $B_{z}^{\text{part}}$ is a particular solution to the differential
equation (\ref{eq:DiffEqBz(zeta)}). The traditional procedure is
to make an educated guess for the solution by choosing a form very
similar to the non-homogeneity factor. In effect, we take $B_{z}^{\text{part}}=B_{\text{eff}}$.
Therefore,
\begin{equation}
B_{z}\left(\zeta\right)=c_{3}I_{0}\left(\zeta\right)+c_{4}K_{0}\left(\zeta\right)+B_{\text{eff}}\qquad\left(\zeta\leqslant\frac{R}{a}\right)\,,\label{eq:SolDiffEqModBesselIn}
\end{equation}
which should be valid for $r<R$, i.e. $a\zeta<R$, including at the
axis of the solenoid where $r=\zeta=0$. At this point, $K_{0}\left(\zeta\right)$
diverges. In fact \cite{Abramowitz},
\begin{equation}
I_{0}\left(z\right)\sim1+\frac{\left(\frac{1}{4}z^{2}\right)}{1!}\Rightarrow\lim_{z\rightarrow0}I_{0}\left(z\right)\rightarrow1\,,\qquad\text{and}\qquad K_{0}\left(z\right)\sim-\left\{ \ln z+\gamma\right\} I_{0}\left(z\right)\Rightarrow\lim_{z\rightarrow0}K_{0}\left(z\right)=\infty\,.\label{eq:I0_K0_smallz}
\end{equation}
where $\gamma\simeq0.57722$ is the Euler-Mascheroni constant. In
order to avoid unphysical $B_{z}$'s divergences we must take: $c_{4}=0$.
This imposition leads to:
\begin{equation}
B_{z}\left(\zeta\right)=c_{3}I_{0}\left(\zeta\right)+B_{\text{eff}}\qquad\left(\zeta\leqslant\frac{R}{a}\right)\,.\label{eq:Bz(c3)in}
\end{equation}

The last step in determining generalized electrodynamics magnetic
field $\mathbf{B}$ is to glue the solutions (\ref{eq:Bz(c2)out})
and (\ref{eq:Bz(c3)in}) at the solenoid's surface. By demanding continuity
of $B_{z}\left(\zeta\right)$ at $\zeta=R/a$, one gets: $c_{2}K_{0}\left(R/a\right)=c_{3}I_{0}\left(R/a\right)+B_{\text{eff}}$.
The imposition of continuity for the derivative of the field, $dB_{z}\left(\zeta\right)/d\zeta$
at $\zeta=R/a$, reads: $-c_{2}K_{1}\left(R/a\right)=c_{3}I_{1}\left(R/a\right)$.
Combining these equation enables us to determine both $c_{2}$ and
$c_{3}$. The calculation is facilitated by utilizing the Wronskian
of $I_{0}$ and $K_{0}$ \cite{Abramowitz}, namely $W\left\{ K_{0}\left(\zeta\right),I_{0}\left(\zeta\right)\right\} =I_{0}\left(\zeta\right)K_{1}\left(\zeta\right)+I_{1}\left(\zeta\right)K_{0}\left(\zeta\right)=1/\zeta$.
The result is the final form of the GE magnetic field for a long,
thin solenoid:
\begin{equation}
B_{z}\left(r\right)=\begin{cases}
\left(\mu_{0}nI\right)\left(R/a\right)I_{1}\left(R/a\right)K_{0}\left(r/a\right)\,, & r>R\\
\left(\mu_{0}nI\right)\left[1-\left(R/a\right)K_{1}\left(R/a\right)I_{0}\left(r/a\right)\right]\,, & r\leqslant R
\end{cases},\label{eq:Bz(s)}
\end{equation}
where we have used (\ref{eq:Beff_solenoid}) and (\ref{eq:zeta}). 

The general solution (\ref{eq:Bz(s)}) can be specified in terms of
elementary functions by recalling that GE coupling constant $a$ should
be small inasmuch the Podolsky contribution is vastly subdominant
with respect to Maxwell's electrodynamics. Therefore, $R\gg a$ and
we may use the asymptotic forms $I_{1}\left(\zeta\right)\simeq e^{\zeta}/\sqrt{2\pi\zeta}$
and $K_{1}\left(\zeta\right)\simeq e^{-\zeta}/\sqrt{\pi/2\zeta}$
for large values of the argument \cite{Abramowitz} to write:
\begin{equation}
B_{z}\left(r\right)=\begin{cases}
\left(\mu_{0}nI\right)\frac{1}{2}\left(\frac{R}{r}\right)^{1/2}e^{-\left(r-R\right)/a}\,, & r>R\gg a\\
\left(\mu_{0}nI\right)\left[1-\left(\frac{\pi}{2}\frac{R}{a}\right)^{1/2}e^{-R/a}I_{0}\left(r/a\right)\right]\,, & r\leqslant R
\end{cases}\label{eq:Bz(s)R_gg_a}
\end{equation}
In the first line, $r>R\gg a$; consequently, we were allowed to approximate
$K_{0}\left(r/a\right)$ by its functional form in Eq. (\ref{eq:I0_K0_largez}). 

\begin{figure}
\begin{center}

\includegraphics[scale=0.5]{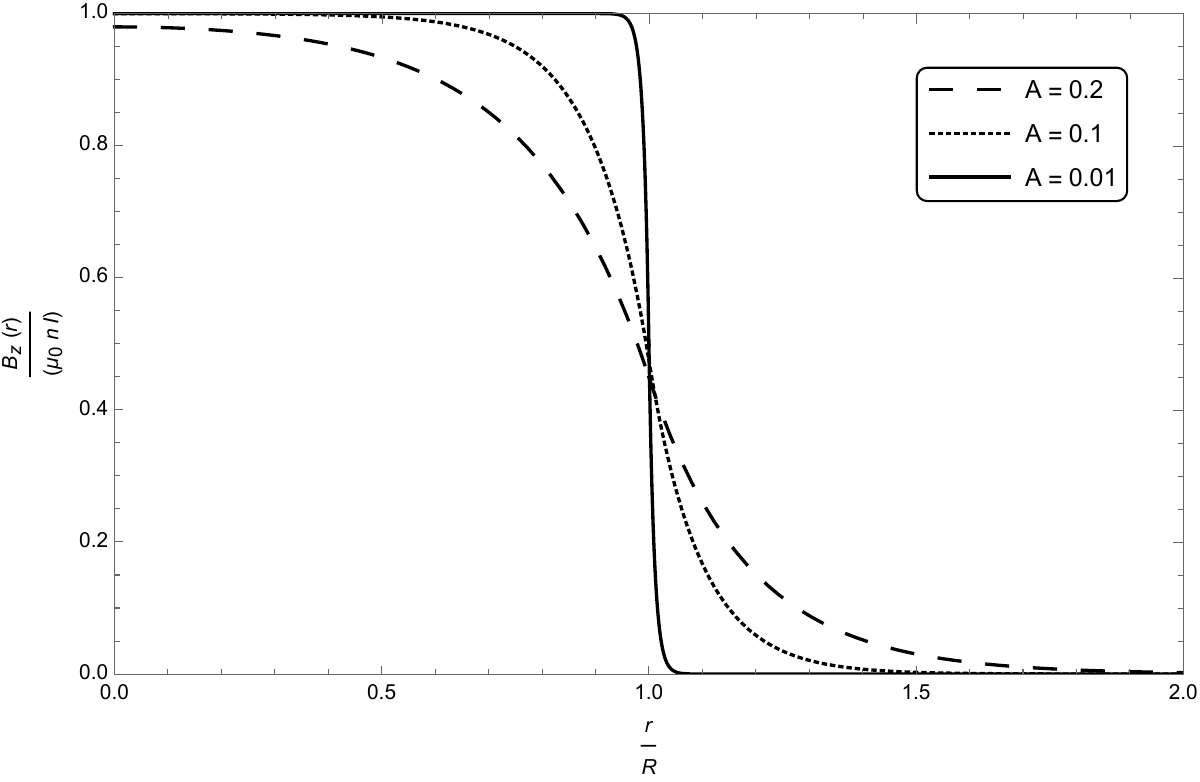}

\end{center}

\caption{Behavior of the magnetic field produced by a long-thin solenoid in
the context of Generalized Electrodynamics as a function of the axial
distance $r$ from the solenoid's axis. The horizontal axis presents
the radial distance $r$ from the solenoid's axis normalized by its
radius $R$. The vertical axis exhibits the magnitude of the field
$\mathbf{B}$ in units of $B_{\text{in}}^{\left(\text{M}\right)}=\left(\mu_{0}nI\right)$
the magnitude of $\mathbf{B}$ inside the solenoid from Maxwell's
electromagnetism. Each curve corresponds to a different choice of
value for Podolsky's parameter $a$ in units of the solenoid's radius
$R$, i.e. $A=a/R$. The smaller the value of $A$, the closer the
behavior of $\mathbf{B}$ to Maxwell's predictions. }
\label{fig:B-Field}
\end{figure}

Also, $I_{0}\left(r/a\right)$ can be approximated by Eq. (\ref{eq:I0_K0_largez})
as long as $r\gg a$. This should be the case for all points except
those very close to the solenoid's axis, where $\lim_{r\rightarrow0}I_{0}\left(r/a\right)=1$
and $B_{z}\simeq\mu_{0}nI$. Hence,
\begin{equation}
B_{z}\left(r\right)\simeq\begin{cases}
\left(\mu_{0}nI\right)\frac{1}{2}\left(\frac{R}{r}\right)^{1/2}e^{-\left(r-R\right)/a}\,, & r>R\gg a\\
\left(\mu_{0}nI\right)\left[1-\frac{1}{2}\left(\frac{R}{r}\right)^{1/2}e^{-\left(R-r\right)/a}\right]\,, & a\ll r\leqslant R
\end{cases}\label{eq:Bz(s)s_gg_a}
\end{equation}
This form of $B_{z}\left(s\right)$ makes it explicit the continuity
at $r=R$. In fact, both lines of (\ref{eq:Bz(s)s_gg_a}) give $\left(\mu_{0}nI\right)/2$
at the solenoid's surface. This is half the value predicted by Maxwell
electrodynamics: it is simply an artifact of the approximations used
from (\ref{eq:Bz(s)}) to (\ref{eq:Bz(s)s_gg_a}). The approximations
select the medium point between the discontinuous Maxwellian magnetic
field, whose value $\left(\mu_{0}nI\right)$ within the solenoid drops
down to zero precisely at $r=R$ and remains null outside the solenoid.
The complete non-approximate solution (\ref{eq:Bz(s)}) exhibits no
such a feature, as shown in the plots of Fig.\ref{fig:B-Field}. $B_{z}\left(r\right)$
decreases smoothly as $r$ increases; it begins with the value $\left(\mu_{0}nI\right)$
at the solenoid's axis $\left(r=0\right)$, transits continuously
through the solenoids surface and approaches zero as $r\rightarrow\infty$.
The precise value of $B_{z}$ at the solenoid's wall depends on the
value of GE coupling constant: the smaller the $a$ the closer the
behaviour of $B_{z}$ to Maxwell's predictions.

\subsection{Vector potential for a long solenoid in GE\label{subsec:A-Potential}}

GE is $U$(1) gauge-invariant \cite{Cuzinatto2007}.\footnote{Actually, GE is \emph{the only} linear generalization of Maxwell Electrodynamics
containing higher-order derivatives of the fields $\mathbf{E}$ and
$\mathbf{B}$ that manifests gauge invariance \cite{Cuzinatto2007}.} Consequently, a vector potential $\mathbf{A}$ respecting the usual
relation to the magnetic field may be introduced:
\begin{equation}
\mathbf{B}=\nabla\times\mathbf{A}\label{eq:rotA}
\end{equation}
Symmetry arguments stated in Section \ref{subsec:B-Field} led us
to conclude that $\mathbf{B}=B_{z}\left(r\right)\hat{\mathbf{z}}$
--- cf. Eq. (\ref{eq:SymmetryOnB}). From this restriction and the
right-hand rule, it follows:
\begin{equation}
\mathbf{A}=A_{\phi}\left(r\right)\hat{\phi}\qquad\left(\text{symmetry}\right)\label{eq:SymmetryOnA}
\end{equation}
 $\left(A_{r}=A_{z}=0\right)$. The curl of $\mathbf{A}$ in cylindrical
coordinates \cite{GriffithsEletro} then reads:
\[
\frac{1}{r}\frac{d}{dr}\left(rA_{\phi}\right)=B_{z}\left(r\right)
\]
i.e.
\begin{equation}
\left(rA_{\phi}\right)=\int rB_{z}\left(r\right)dr\,,\label{eq:sAphi}
\end{equation}
up to an integration constant (recall that $A_{\phi}$ is a function
of $r$ only); this constant will be added later on. $B_{z}\left(r\right)$
is given by Eq. (\ref{eq:Bz(s)}) and splits naturally in two different
functions according to the interval of values assumed by $r$.

Outside the solenoid, the integral in (\ref{eq:sAphi}) reads:
\begin{align}
\int rB_{z}\left(r\right)dr & =\left(\mu_{0}nI\right)\left(R/a\right)I_{1}\left(R/a\right)\int rK_{0}\left(r/a\right)dr\nonumber \\
 & =-\left(\mu_{0}nI\right)a^{2}\left(R/a\right)\left(r/a\right)I_{1}\left(R/a\right)K_{1}\left(r/a\right)\qquad\left(r>R\right)\,,\label{eq:intsBzds(K1)}
\end{align}
The modified Bessel functions recurrence relations were used in the
last step. In particular, the second equation in paragraph 9.6.26
of Ref. \cite{Abramowitz} leads to $\frac{d}{dz}\left[zK_{1}\left(z\right)\right]=-zK_{0}\left(z\right)$.
Inserting (\ref{eq:intsBzds(K1)}) into (\ref{eq:sAphi}):
\begin{equation}
A_{\phi}=-\left(\mu_{0}nI\right)a\left(\frac{R}{a}\right)I_{1}\left(R/a\right)K_{1}\left(r/a\right)+\left(\frac{a}{r}\right)A_{\text{out}}\qquad\left(r>R\right)\,,\label{eq:Aphi(Aout)}
\end{equation}
where is $\left(aA_{\text{out}}\right)$ the integration constant
omitted in (\ref{eq:sAphi}). As a remark, notice the following. The
asymptotic form of $K_{1}\left(z\right)$ for large $z$ is obtained
from Eq. 9.7.2 in Ref. \cite{Abramowitz}: $K_{1}\left(z\right)\sim\sqrt{\frac{\pi}{2z}}e^{-z}\Rightarrow\lim_{z\rightarrow\infty}K_{1}\left(z\right)=0$,
so that $\lim_{r\rightarrow\infty}A_{\phi}\left(r\right)=0$. This
shows that (\ref{eq:Aphi(Aout)}) is consistent with what is expected
from ordinary Maxwellian electrodynamics \cite{GriffithsEletro}.

Inside the solenoid,
\begin{align}
\int rB_{z}\left(r\right)dr & =\left(\mu_{0}nI\right)\left[\int rdr-\left(R/a\right)K_{1}\left(R/a\right)\int rI_{0}\left(r/a\right)dr\right]\nonumber \\
 & =\left(\mu_{0}nI\right)a^{2}\left[\left(r/a\right)^{2}/2-\left(R/a\right)\left(r/a\right)K_{1}\left(R/a\right)I_{1}\left(r/a\right)\right]\qquad\left(r\leqslant R\right)\,.\label{eq:intsBzds(I1)}
\end{align}
Here, we have utilized Eq. (\ref{eq:Bz(s)}) and the fact that $\frac{d}{dz}\left[zI_{1}\left(z\right)\right]=zI_{0}\left(z\right)$.
This last statement is a consequence of the properties listed in the
paragraph 9.6.26 of Ref. \cite{Abramowitz}. Due to (\ref{eq:intsBzds(I1)}),
Eq. (\ref{eq:sAphi}) gives:
\begin{equation}
A_{\phi}=\left(\mu_{0}nI\right)a\left[\frac{1}{2}\left(\frac{r}{a}\right)-\left(\frac{R}{a}\right)K_{1}\left(R/a\right)I_{1}\left(r/a\right)\right]+\left(\frac{a}{r}\right)A_{\text{in}}\qquad\left(r\leqslant R\right)\,.\label{eq:Aphi(Ain)}
\end{equation}
The integration constant $\left(aA_{\text{in}}\right)$ is determined
by demanding consistency with the Maxwellian case --- denoted by
a superscript (M),
\begin{equation}
A_{\phi}^{(\text{M})}=\left(\mu_{0}nI\right)\frac{r}{2}\qquad\left(r\leqslant R\right)\,.\label{eq:Aphi(M)in}
\end{equation}
{[}See e.g. Ref. \cite{GriffithsEletro}, Eq. (5.70).{]} In effect,
Eq. 9.6.10 in Ref. \cite{Abramowitz} gives the asymptotic form of
$I_{1}\left(z\right)$ for small $z$: $I_{1}\left(z\right)\sim\frac{1}{2\Gamma\left(2\right)}z\Rightarrow\lim_{z\rightarrow0}I_{1}\left(z\right)=0$.
Therefore, $\lim_{r\rightarrow0}A_{\phi}\left(r\right)=\lim_{r\rightarrow0}\left(\frac{a}{r}\right)A_{\text{in}}\rightarrow\infty$.
In order to avoid this divergence in (\ref{eq:Aphi(Ain)}) and make
it completely consistent with (\ref{eq:Aphi(M)in}), we must impose
\begin{equation}
A_{\text{in}}=0\,.\label{eq:Ain_null}
\end{equation}

Furthermore, we demand continuity of $A_{\phi}$ at the surface of
the solenoid. Equating (\ref{eq:Aphi(Aout)}) and (\ref{eq:Aphi(Ain)})
at $r=R$, results:
\begin{equation}
A_{\text{out}}=\left(\mu_{0}nI\right)\frac{a}{2}\left(\frac{R}{a}\right)^{2}\label{eq:Aout}
\end{equation}
Eqs. (\ref{eq:Ain_null}) and (\ref{eq:Aout}) put us in position
to write down (\ref{eq:Aphi(Aout)}) and (\ref{eq:Aphi(Ain)}) as
the complete solution for $A_{\phi}$ in the context of GE: 
\begin{equation}
A_{\phi}\left(r\right)=\begin{cases}
\left(\mu_{0}nI\right)R\left[\frac{1}{2}\left(\frac{R}{r}\right)-I_{1}\left(R/a\right)K_{1}\left(r/a\right)\right]\qquad\left(r>R\right)\\
\left(\mu_{0}nI\right)R\left[\frac{1}{2}\left(\frac{r}{R}\right)-K_{1}\left(R/a\right)I_{1}\left(r/a\right)\right]\qquad\left(r\leqslant R\right)
\end{cases}\,.\label{eq:Aphi(s)}
\end{equation}
The first term in line one is precisely the vector potential outside
the solenoid in Maxwell's case \cite{GriffithsEletro}:
\begin{equation}
A_{\phi}^{(\text{M})}=\left(\mu_{0}nI\right)\frac{1}{2}\frac{R^{2}}{r}\qquad\left(r>R\right)\,.\label{eq:Aphi(M)out}
\end{equation}

\begin{figure}
\begin{center}

\includegraphics[scale=0.5]{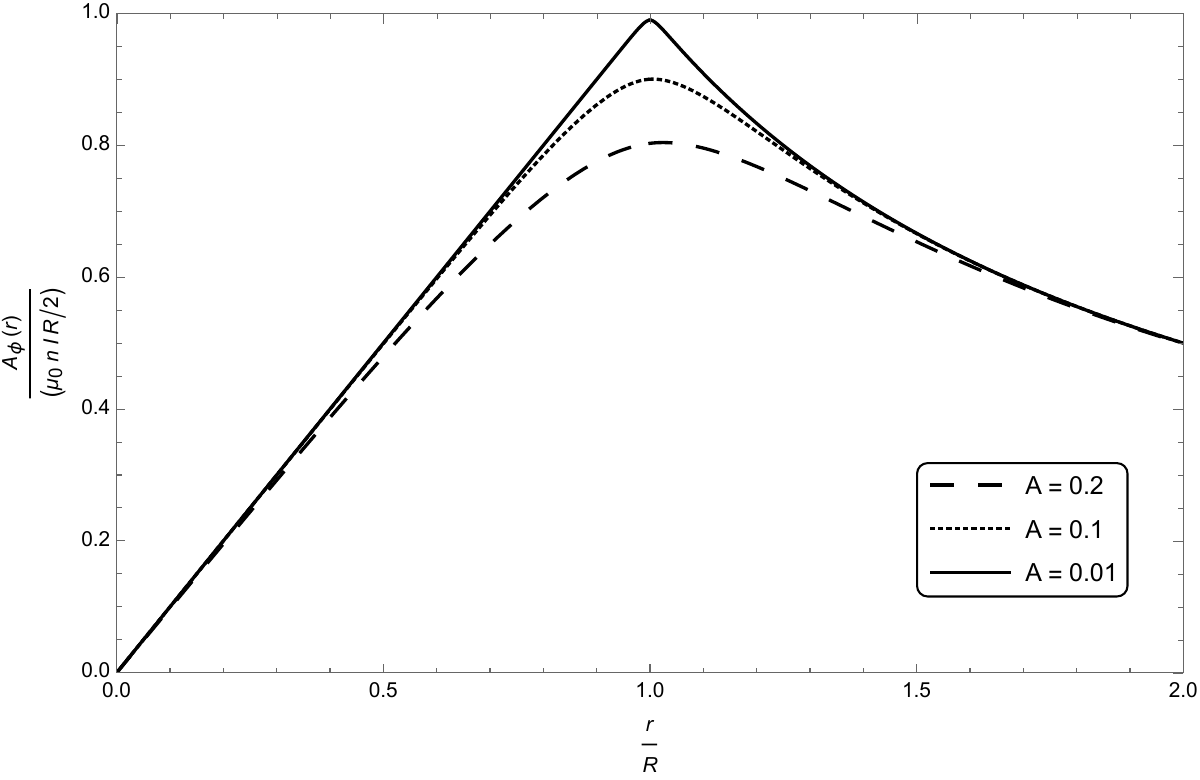}

\end{center}

\caption{Behavior of the vector potential produced by a long-thin solenoid
in the context of Generalized Electrodynamics. The horizontal axis
is the distance from the solenoid's axis $r$normalized by the solenoid's
radius $R$. The vertical axis shows the vector potential $A_{\phi}\left(r\right)$
weighted by its value at the solenoid's surface $\left(\mu_{0}nI\right)\left(R/2\right)$.
The definition $A=a/R$ is used here.}
\label{fig:A-Potential}
\end{figure}

Just like in the magnetic field case, one can take into account that
$R\gg a$ and approximate $I_{1}\left(R/a\right)$ and $K_{1}\left(R/a\right)$
in terms of elementary functions. Eq. (\ref{eq:Aphi(s)}) then reads:
\begin{equation}
A_{\phi}\left(r\right)=\begin{cases}
\left(\mu_{0}nI\right)\frac{1}{2}\frac{R^{2}}{r}\left[1-\left(\frac{a}{R}\right)\left(\frac{r}{R}\right)^{1/2}e^{-\left(r-R\right)/a}\right]\qquad\left(r>R\gg a\right)\\
\left(\mu_{0}nI\right)\frac{1}{2}\frac{R^{2}}{r}\left[\left(\frac{r}{R}\right)^{2}-\sqrt{2\pi}\left(\frac{a}{R}\right)^{1/2}\left(\frac{r}{R}\right)e^{-R/a}I_{1}\left(r/a\right)\right]\qquad\left(s\leqslant R\right)
\end{cases}\label{eq:Aphi(s)R_gg_a}
\end{equation}
Wherever $r$ is not too close to zero (axis of the solenoid), it
is reasonable to take one more approximation, namely $r\gg a.$ In
this case,
\begin{equation}
A_{\phi}\left(r\right)=\begin{cases}
\left(\mu_{0}nI\right)\frac{1}{2}\frac{R^{2}}{r}\left[1-\left(\frac{a}{R}\right)\left(\frac{r}{R}\right)^{1/2}e^{-\left(r-R\right)/a}\right]\qquad\left(r>R\gg a\right)\\
\left(\mu_{0}nI\right)\frac{1}{2}\frac{R^{2}}{r}\left[\left(\frac{r}{R}\right)^{2}-\left(\frac{a}{R}\right)\left(\frac{r}{R}\right)^{1/2}e^{-\left(R-r\right)/a}\right]\qquad\left(a\ll r\leqslant R\right)
\end{cases}\label{eq:Aphi(s)s_gg_a}
\end{equation}
This equation makes it clear that both the vector potential $A_{\phi}\left(s\right)$
and its derivative is continuous at the solenoid's surface---see
also Fig. \ref{fig:A-Potential} produced with Eq. (\ref{eq:Aphi(s)}).

\subsection{Magnetic Aharonov-Bohm phase shift in GE\label{sec:Magnetic-AB-phase-shift}}

\begin{figure}
\begin{center}

\includegraphics[scale=0.1]{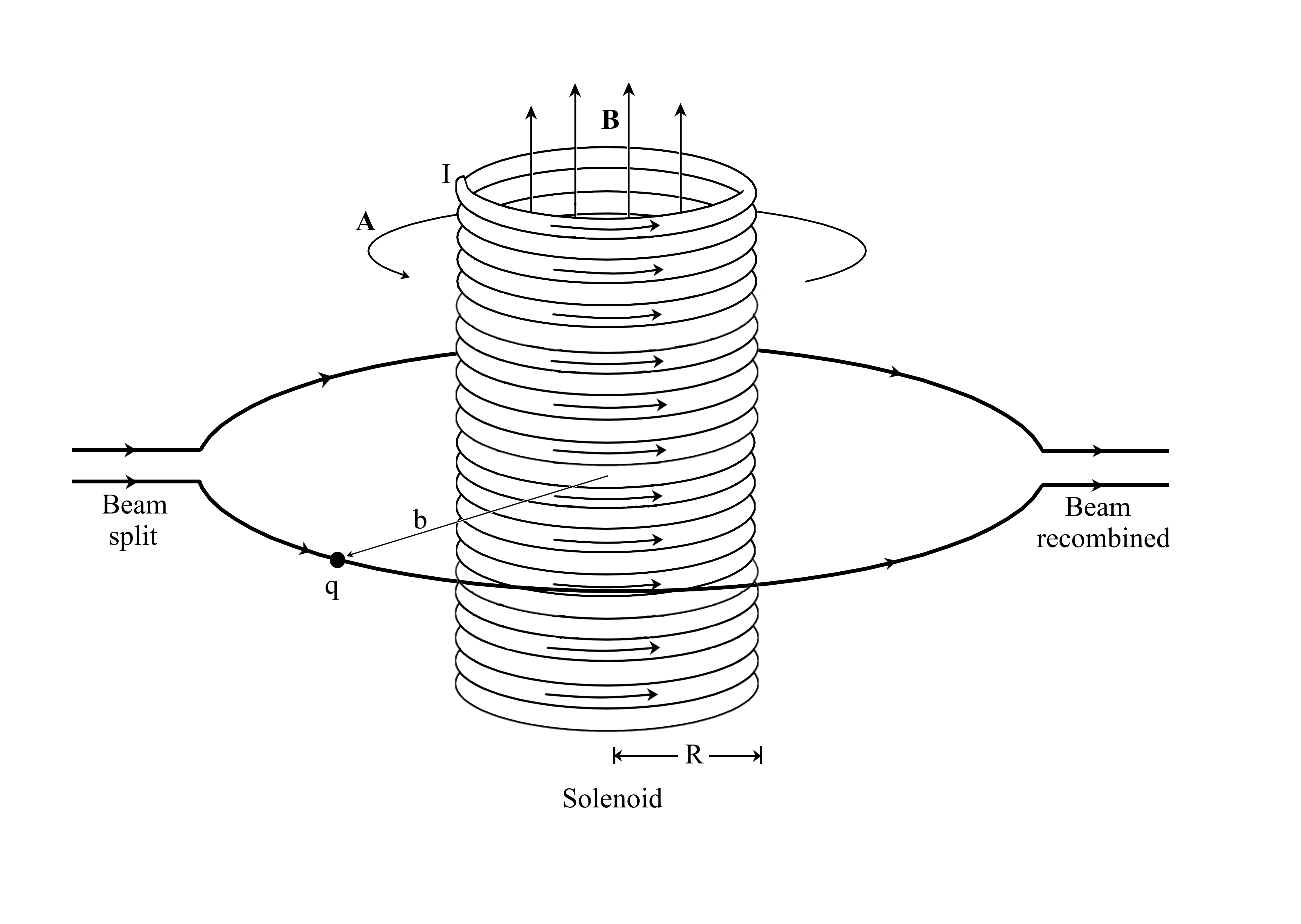}

\end{center}

\caption{The beam of electrons splits before circling the solenoid. The trajectory
of each half of the beam is a semi-circle. Both branches from a ring-shaped
path. }
\label{fig:ABsetup}
\end{figure}

Consider the experimental configuration in Fig. \ref{fig:ABsetup}(a).
Therein, an electron beam splits off on the outside vicinity of a
long thin solenoid; the two resulting branches of the original beam
wrap around the solenoid in a ring shape path whose plane is perpendicular
to the solenoid's axis. The two halves of the beam reunite in a single
beam before hitting the detector. The detector measures the magnetic
Aharonov-Bohm phase shift $\Delta g$. 

The recipe for calculating AB phase $g$ is provided by the path integral
formulation in Chapter 2 of Ref. \cite{Felsager}. Therein, we see
that the interaction of the particle with the electromagnetic field
is accounted for through the action integral
\begin{equation}
S_{I}=q\int A_{\alpha}dx^{\alpha}\,.\label{eq:S_I}
\end{equation}
(Einstein's sum convention admited). $A_{\alpha}$ is the four-vector
potential.\footnote{$A^{\alpha}=\left(A^{0},A^{1},A^{2},A^{3}\right)=\left(-\varphi/c,A_{r},A_{\theta},A_{\phi}\right)$
in cylindrical coordinates.} We also learn from Ref. \cite{Felsager} that the action $S_{I}$
contributes as an additional phase factor $\Phi_{I}$
\begin{equation}
\Phi_{I}=\exp\left[\frac{i}{\hbar}S_{I}\right]\label{eq:Phi_I}
\end{equation}
to the wavefunction $\Psi$. Eq. (\ref{eq:S_I}) follows from the
Lagrangian formalism for particles together with the equation for
the Lorentz force of electromagnetism. Now, the Lorentz force in not
derived from Maxwell equation of electrodynamics. It is an additional
piece of the theory. In GE, Maxwell equations are generalized to contain
higher-order derivatives of the field strengths. However, the Lorentz
force equation is unchanged in GE and so is the Lagrangian formalism
involving the action integral. Therefore, Eqs. (\ref{eq:S_I}) and
(\ref{eq:Phi_I}) should apply to GE's case. In fact, we have $g=S_{I}/\hbar$,
i.e.
\[
g=\frac{q}{\hbar}\int_{\Gamma}A_{\alpha}dx^{\alpha}
\]
where $\Gamma$ is the integration path. In Aharonov-Bohm experimental
sets, we measure the phase diference $\Delta g$ in closed paths,
such as that represented in Fig. \ref{fig:ABsetup}. In effect, we
are interested in
\begin{equation}
\Delta g=g_{2}-g_{1}=\frac{q}{\hbar}\left(\int_{\Gamma_{2}}A_{\alpha}dx^{\alpha}-\int_{\Gamma_{1}}A_{\alpha}dx^{\alpha}\right)\,,\label{eq:Deltag(Gamma)}
\end{equation}
where is the $\Gamma_{2}$ ($\Gamma_{1}$) is the branch on the right
(left) of the solenoid with respect to the direction of the beam in
Fig. \ref{fig:ABsetup}(a). The negative sign on the right hand side
of Eq. (\ref{eq:Deltag(Gamma)}) suggests that we integrate backwards
along $\Gamma_{1}$ (whilst we integrate forward along $\Gamma_{2}$).
This results in the loop integral:
\begin{equation}
\Delta g=\frac{q}{\hbar}\oint A_{\alpha}dx^{\alpha}\label{eq:Deltag}
\end{equation}
This is the expression we are seeking for calculating the AB effect
in the context of GE. 

\begin{figure}
\begin{center}

\includegraphics[scale=0.5]{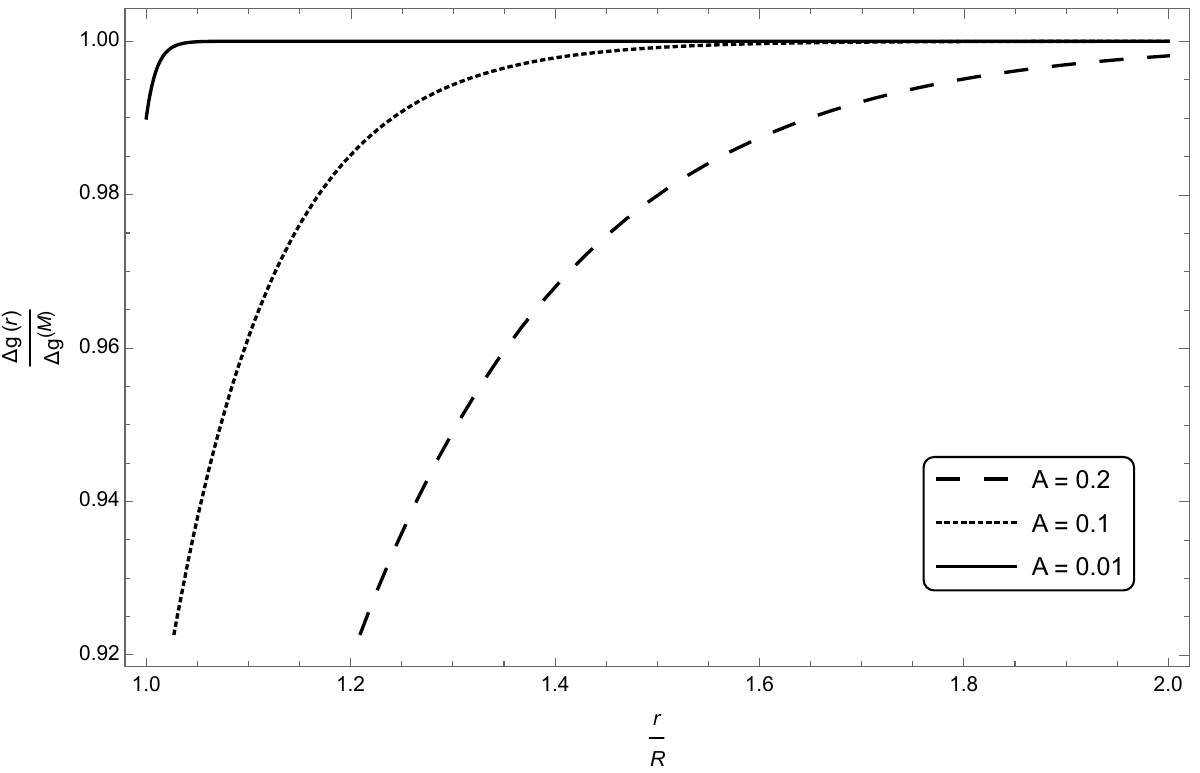}

\end{center}

\caption{Relative phase shift in the magnetic Aharonov-Bohm effect in the context
of Generalized Electrodynamics. The correction $\delta g\left(r\right)$
changes Maxwellian prediction in the exterior vicinity of the solenoid.
The more increases the distance to the solenoid the smaller the effectiveness
of GE on the phase shift.}
\label{fig:Phase-Shift}
\end{figure}

Therefore:
\begin{equation}
\Delta g=\frac{q}{\hbar}\oint A_{0}dx^{0}+\frac{q}{\hbar}\oint A_{j}dx^{j}\label{eq:Deltag_AB}
\end{equation}
where the three-vector $A_{j}$ is $\mathbf{A}=A_{\phi}\left(r\right)\hat{\phi}$
due to (\ref{eq:SymmetryOnA}) and $A_{0}=-A^{0}=-\varphi/c=0$ because
there is no electric potential in the pure magnetic AB effect. Then,
\begin{equation}
\Delta g=\frac{q}{\hbar}\oint\mathbf{A}\cdot d\mathbf{l}\label{eq:Deltag_AB_Mag}
\end{equation}
with
\begin{equation}
d\mathbf{l}=dr\,\hat{r}+rd\phi\,\hat{\phi}+dz\,\hat{z}\label{eq:dl_cylindrical}
\end{equation}
is the line element in cylindrical coordinates. Let us use the first
line of (\ref{eq:Aphi(s)R_gg_a}) into (\ref{eq:Deltag_AB_Mag}):
\begin{align*}
\Delta g & =\frac{q}{\hbar}\oint\left[A_{\phi}\left(r\right)\hat{\phi}\right]\cdot\left[dr\,\hat{r}+rd\phi\,\hat{\phi}+dz\,\hat{z}\right]=\frac{q}{\hbar}\oint A_{\phi}\left(r\right)rd\phi\\
 & =\frac{q}{\hbar}A_{\phi}\left(r\right)r\oint d\phi=\frac{q}{\hbar}\left(\mu_{0}nI\right)\frac{1}{2}\frac{R^{2}}{r}\left[1-\left(\frac{a}{R}\right)\left(\frac{r}{R}\right)^{1/2}e^{-\left(r-R\right)/a}\right]r\oint d\phi\qquad\left(r>R\right)\,.
\end{align*}
Consider once again the setup in Fig. \ref{fig:ABsetup}(a). One can
orient the $y$-axis in the direction of the outgoing beam; so that
the $x$-axis points outward the page. We call right (left) branch
the part of the beam circling the solenoid in the positive (negative)
region of the $x$-axis. Then, the angle goes from $-\pi/2$ to $\pi/2$
in the right branch of the beam and from $3\pi/2$ to $\pi/2$ in
the negative branch in the clockwise direction. Then, 
\[
\oint d\phi=\int_{-\pi/2}^{\pi/2}d\phi+\int_{3\pi/2}^{\pi/2}d\left(-\phi\right)=\left.\phi\right|_{-\pi/2}^{\pi/2}+\left.\phi\right|_{\pi/2}^{3\pi/2}=\left[\frac{\pi}{2}-\left(-\frac{\pi}{2}\right)\right]+\left(\frac{3\pi}{2}-\frac{\pi}{2}\right)=2\pi\,,
\]
as it should be in any case. Hence,
\[
\Delta g=\frac{q}{\hbar}\left(\mu_{0}nI\right)\frac{1}{2}\frac{R^{2}}{r}\left[1-\left(\frac{a}{R}\right)\left(\frac{r}{R}\right)^{1/2}e^{-\left(r-R\right)/a}\right]r\left(2\pi\right)\,,
\]
which we will write in terms of the absolute value of the magnetic
field inside the solenoid in Maxwell electrodynamics:
\begin{equation}
B_{z}^{\left(M\right)}=\mu_{0}nI\qquad\left(r\leqslant R\right)\label{eq:Bz(M)in}
\end{equation}
and the solenoid's cross section 
\begin{equation}
S=\pi R^{2}\,;\label{eq:S_cross_section}
\end{equation}
i.e. 
\begin{align*}
\Delta g & =\frac{q}{\hbar}B_{z}^{\left(M\right)}\left(\pi R^{2}\right)\left[1-\left(\frac{a}{R}\right)\left(\frac{r}{R}\right)^{1/2}e^{-\left(r-R\right)/a}\right]\\
 & =\frac{q}{\hbar}\left[B_{z}^{\left(M\right)}S\right]\left[1-\left(\frac{a}{R}\right)\left(\frac{r}{R}\right)^{1/2}e^{-\left(r-R\right)/a}\right]\,.
\end{align*}
Now, 
\begin{equation}
\Phi^{\left(M\right)}=B_{z}^{\left(M\right)}S\label{eq:Phi(M)}
\end{equation}
is the flux of the magnetic field through the solenoid as predicted
by Maxwell electrodynamics. Thus,
\begin{equation}
\Delta g=\frac{q\Phi^{\left(M\right)}}{\hbar}\left[1-\left(\frac{a}{R}\right)\left(\frac{r}{R}\right)^{1/2}e^{-\left(r-R\right)/a}\right]\qquad\left(r>R\right)\,.\label{eq:PhaseShift_Podolsky}
\end{equation}
The term in front of the square brackets is precisely the phase shift
expected in Maxwellian electrodynamics---see Eq. (\ref{eq:Deltag(M)}):
\begin{equation}
\Delta g^{\left(M\right)}=\frac{q\Phi^{\left(M\right)}}{\hbar}\,.\label{eq:Deltag(M)}
\end{equation}
Therefore,
\begin{equation}
\frac{\Delta g}{\Delta g^{\left(M\right)}}=\left[1-\delta g\left(r\right)\right]\,,\label{eq:Deltag(Deltag(M))}
\end{equation}
with the phase shift displacement due to GE being:
\begin{equation}
\delta g\left(r\right)\equiv\left(\frac{a}{R}\right)\left(\frac{r}{R}\right)^{1/2}e^{-\left(r-R\right)/a}\qquad\left(r>R\right)\,.\label{eq:deltag(P)}
\end{equation}
Eq. (\ref{eq:PhaseShift_Podolsky})--- or Eq. (\ref{eq:Deltag(Deltag(M))})---give
the phase shift $\Delta g$ for the magnetic Aharonov-Bohm effect
in the context of Podolsky electrodynamics. The radial dependence of
$\Delta g$ is a novelty of the AB effect in GE: in Maxwell electrodynamics
is a constant, cf. Eq. (\ref{eq:Deltag(M)}). Inasmuch the GE typical
effectiveness range $a$ is a small constant, $a\ll R$, the phase
shift deviation $\delta g$ is a tiny amount. Fig. \ref{fig:Phase-Shift}
shows the plot of $\Delta g/\Delta g^{(M)}$ as a function of the
radial distance $s$.

\vspace{0.5cm}

The perturbation $\delta g\left(r\right)$ due to GE is exponentially
suppressed with the increasing distance $r>R$ from the exterior of
the solenoid. The maximum value for the phase shift displacement is
$\delta g_{\text{max}}\left(R\right)=a/R$ and occurs at $r=R$, the
outside surface of solenoid. Therefore, the best case scenario for
detection of the Aharonov-Bohm effect in Generalized Electrodynamics
is to produce a beam splitting in which the right and left branches
circle around the solenoid almost touching its exterior surface. In
this case, 
\[
r=R+\delta r\qquad\left(\delta r\ll R\right)\,.
\]
Actually, one should achieve $\delta r\sim a$ to access the distance
scales where GE is effective. Whether this is experimentally possible
is still to be determined. Regardless, let us write
\[
\delta r=\epsilon a\qquad\left(\epsilon\lesssim1\right)\,,
\]
and write down $\delta g$ for a beam skimming through the solenoid's
external surface:
\begin{align*}
\delta g^{\left(\text{flyby}\right)} & \equiv\delta g\left(r\sim R\right)=\left(\frac{a}{R}\right)\left(\frac{R+\epsilon a}{R}\right)^{1/2}e^{-\left(R+\epsilon a-R\right)/a}\\
 & =\left(\frac{a}{R}\right)\left(1+\epsilon\frac{a}{R}\right)^{1/2}e^{-\epsilon}\simeq\left(\frac{a}{R}\right)\left(1+\frac{1}{2}\epsilon\frac{a}{R}\right)\left(1-\epsilon\right)
\end{align*}
If we keep up to first order terms in $\epsilon$, we have
\begin{equation}
\delta g^{\left(\text{flyby}\right)}\simeq\left(\frac{a}{R}\right)\left[1-\epsilon\left(1-\frac{1}{2}\frac{a}{R}\right)\right]\simeq\left(\frac{a}{R}\right)\left(1-\epsilon\right)\label{eq:deltag_flyby}
\end{equation}
This is the quantity that should be aimed at experimentally.

\section{Electric AB effect\label{sec:Electric-AB-effect}}

\begin{figure}
\begin{center}

\includegraphics[scale=0.08]{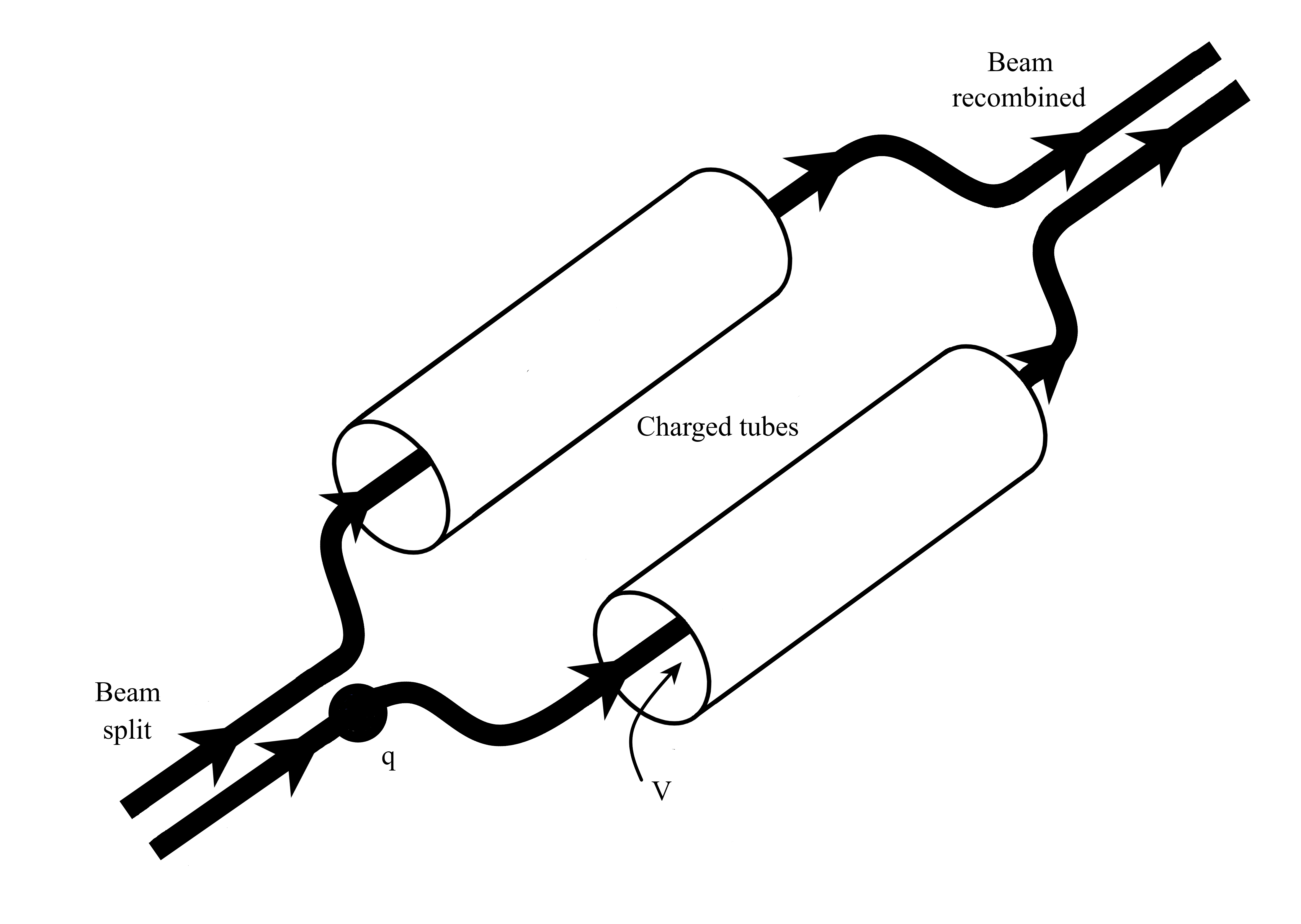}

\end{center}

\caption{The beam of electrons splits in two branches that enter two long cylinders
of charges. The recombined branches hit the detector located in the
upper-right corner where the AB phase shift is measured. }
\label{fig:ABsetup_Electric}
\end{figure}

The electric version of AB effect is obtained in an experimental configuration
schematically represented in Fig. \ref{fig:ABsetup_Electric}. The
problem is the dual of the magnetic AB effect---see Fig. \ref{fig:ABsetup}.
Now, the electric beam splits in two hemispherical branches that pass
through two long tubes of charges before recombining and finally hitting
the detector. In Maxwell electrodynamics there is no electric field
inside the cylinders of charge, so that no effect would be expected
upon the electrons in the beam branches; however, a phase shift is
detected and explained on the basis of the influence of the non-null
electric potential $\varphi$ within the tubes---cf. the first term
in Eq. (\ref{eq:Deltag_AB}) and $A_{0}=-\varphi/c$ . There will
be an analog of the electric AB effect in the context of GE. In order
to start modelling the problem, it is necessary to compute the electric
field $\mathbf{E}$ predicted by GE inside a long tube of static charges.

\subsection{Electric field of a long charged tube in GE \label{subsec:E-field}}

We begin by repeating the strategy of Section \ref{subsec:B-Field}.

Take the Gauss-Podolsky law, i.e. the first equation in (\ref{eq:FieldEqsGE}):
\begin{equation}
\left(1-a^{2}\square\right)\left(\nabla\cdot\mathbf{E}\right)=\frac{1}{\epsilon_{0}}\rho\,.\label{eq:GaussPod}
\end{equation}
For an static uniforme distribution of charges on the surface of a
thin tube of radius $R$, one does not expect $\mathbf{E}$ to depend
on time $t$. Therefore, Eq. (\ref{eq:GaussPod}) reduces to:
\begin{equation}
\nabla\cdot\left[\left(1-a^{2}\nabla^{2}\right)\mathbf{E}\right]=\frac{1}{\epsilon_{0}}\rho\,,\label{eq:GaussPod_static}
\end{equation}
since the Laplacian operator commutes with the divergent of $\mathbf{E}$.
Eq. (\ref{eq:GaussPod_static}) which motivates the definition of
an effective electric field: 
\begin{equation}
\mathbf{E}_{\text{eff}}\equiv\left(1-a^{2}\nabla^{2}\right)\mathbf{E}\,,\label{eq:Eeff}
\end{equation}
that respects Gauss's law in its traditional form: 
\begin{equation}
\nabla\cdot\mathbf{E}_{\text{eff}}=\frac{1}{\epsilon_{0}}\rho\,.\label{eq:GaussEeff}
\end{equation}
The integral form of this equation is: 
\begin{equation}
\ointop_{S}\mathbf{E}_{\text{eff}}\cdot d\mathbf{a}=\frac{1}{\epsilon_{0}}q_{\text{enc}}\label{eq:Gauss}
\end{equation}
where 
\begin{equation}
q_{\text{enc}}=\intop_{V}\rho dV\label{eq:qenc}
\end{equation}
is the charge enclosed by the Gaussian surface $S$ of volume $V$
and surface element $d\mathbf{a}=\mathbf{n}da$ ($\mathbf{n}$ is
an unity vector pointing outward the closed surface).

Eq. (\ref{eq:Gauss}) can be applied to one of the cylinders in Fig.
\ref{fig:ABsetup_Electric}. This is a textbook problem typical of
the course Electromagnetism 101. In fact, take a Gaussian in a shape
of a closed cylinder of radius $r$ and length $l$. The axial symmetry
of the problem imposes $\mathbf{E}_{\text{eff}}=\mathbf{E}_{\text{eff}}\left(r\right)=E_{\text{eff}}\left(r\right)\hat{\mathbf{r}}$.
Then, the left-hand side of (\ref{eq:Gauss}) is: $\ointop_{S}\mathbf{E}_{\text{eff}}\cdot d\mathbf{a}=E_{\text{eff}}\left(r\right)\left(2\pi r\right)\left(l\right)$.
The right-hand side of (\ref{eq:Gauss}) depends on the amount of
charge enclosed by the Gaussian cylinder. If the Gaussian surface
is within the tube of charges, it encloses none of the superficial
charge: $q_{\text{enc}}=0$ for $s<R$. If the radius of the Gaussian
cylinder is greater than the radius $R$ of the tube of charges, then
$q_{\text{enc}}=\sigma\left(2\pi R\right)\left(l\right)$ for $s\geqslant R$,
where we have admitted a uniform surface charge distribution: $\sigma=\text{constant}$.
Equating the previous results yields: 
\begin{equation}
\mathbf{E}_{\text{eff}}\left(r\right)=\begin{cases}
0 & r<R\quad\left(\text{inside the tube}\right)\\
\frac{\sigma}{\epsilon_{0}}\frac{R}{r}\hat{\mathbf{r}}\,, & r\geqslant R\quad\left(\text{outside the tube}\right)
\end{cases}\,.\label{eq:Eeff_tube}
\end{equation}

Plugging this result into Eq. (\ref{eq:Eeff}), leads to the differential
equation for the electric field in the context of GE: 
\begin{equation}
\mathbf{E}-a^{2}\nabla^{2}\mathbf{E}=\begin{cases}
0\,, & r<R\quad\left(\text{inside the tube}\right)\\
\frac{\sigma}{\epsilon_{0}}\frac{R}{r}\hat{\mathbf{r}}\,, & r\geqslant R\quad\left(\text{outside the tube}\right)
\end{cases}\,.\label{eq:DiffEqE(Eeff)}
\end{equation}
In the limit as $a\rightarrow0$, $\mathbf{E}=\mathbf{E}_{\text{eff}}$
and the Maxwellian result is recovered, as it should be. In this case,
we see that the electric field points in the axial direction $\hat{\mathbf{r}}$;
moreover, it depends only on the axial coordinate $r$. We will assume
these features are shared by electric field of GE: 
\begin{equation}
\mathbf{E}=\mathbf{E}\left(r\right)=E_{r}\left(r\right)\hat{\mathbf{r}}\qquad\left(\text{symmetry}\right)\label{eq:SymmetryOnE}
\end{equation}
Due to (\ref{eq:SymmetryOnE}), the functional form of the Laplacian
of the electric field reduces considerably and Eq. (\ref{eq:DiffEqE(Eeff)})
reads:
\begin{equation}
\zeta^{2}\frac{d^{2}E_{r}}{d\zeta^{2}}+\zeta\frac{dE_{r}}{d\zeta}-\zeta^{2}E_{r}=0\,,\qquad\left(\zeta<\frac{R}{a}\right)\label{eq:DiffEqE(zeta)ModBessel}
\end{equation}
and 
\begin{equation}
\zeta^{2}\frac{d^{2}E_{r}}{d\zeta^{2}}+\zeta\frac{dE_{r}}{d\zeta}-\zeta^{2}E_{r}=\left(-\frac{\sigma}{\epsilon_{0}}\frac{R}{a}\right)\zeta\,,\qquad\left(\zeta\geqslant\frac{R}{a}\right)\,.\label{eq:DiffEqE(zeta)ModBesselNonHom}
\end{equation}
Herein, use was make of the dimensionless coordinate $\zeta=r/a$.

In the following we will solve Eqs. (\ref{eq:DiffEqE(zeta)ModBessel})
and (\ref{eq:DiffEqE(zeta)ModBesselNonHom}) one at a time.

By comparing Eq. (\ref{eq:DiffEqE(zeta)ModBessel}) with Eq. 9.6.1
of Ref. \cite{Abramowitz}, we conclude that it as a modified Bessel
equation with $\nu=0$. As such, it is solved by: 
\begin{equation}
E_{r}\left(\zeta\right)=b_{1}I_{0}\left(\zeta\right)+b_{2}K_{0}\left(\zeta\right)\,.\label{eq:SolDiffEqE(zeta)ModBessel}
\end{equation}
The integration constants $\left(b_{1},b_{2}\right)$ are constrained
through imposition of boundary conditions. We demand that $E_{s}\left(\zeta\right)$
is regular at the tube's axis, where $\zeta=0$. From the fact that
$\lim_{z\rightarrow0}K_{0}\left(z\right)\rightarrow\infty$, we must
choose 
\begin{equation}
b_{2}=0\label{eq:b2}
\end{equation}
in order to avoid an unphysical divergent $E_{r}$. The result is:
\begin{equation}
E_{r}\left(\zeta\right)=b_{1}I_{0}\left(\zeta\right)\qquad\left(\zeta<\frac{R}{a}\right)\,.\label{eq:Es(b1)in}
\end{equation}
The constant $b_{1}$ is unconstrained for the time being. Later it
will be fixed by requiring continuity with the electric field outside
the tube of charges. 

This leads us to our next task: to solve the inhomogeneous modified
Bessel equation (\ref{eq:DiffEqE(zeta)ModBesselNonHom}). We try a
solution of the type 
\begin{equation}
E_{r}=E_{r}^{\text{hom}}+E_{r}^{\text{part}}\label{eq:EsHomPlusEsPart}
\end{equation}
where 
\begin{equation}
E_{r}^{\text{hom}}=b_{3}I_{0}\left(\zeta\right)+b_{4}K_{0}\left(\zeta\right)\label{eq:EsHom}
\end{equation}
and, we have, 
\begin{equation}
E_{r}^{\text{part}}=\left(-\frac{\sigma}{\epsilon_{0}}\frac{R}{a}\right)\frac{\pi}{2}L_{0}\left(\zeta\right)\,.\label{eq:EsPart}
\end{equation}
Here $L_{\nu}\left(z\right)=-ie^{-i\nu\pi/2}H_{\nu}\left(z\right)$
is the modified Struve function---see e.g. Ref. \cite{Abramowitz},
Section 12.2. It is defined in term of the Struve function $H_{\nu}\left(z\right)$
which composes the general solution of the differential equation
\begin{equation}
z^{2}\frac{d^{2}w}{dz^{2}}+z\frac{dw}{dz}+\left(z^{2}-\nu^{2}\right)w=\frac{4\left(\frac{1}{2}z\right)^{\nu+1}}{\sqrt{\pi}\Gamma\left(\nu+\frac{1}{2}\right)}\label{eq:StruveDiffEq}
\end{equation}
alongside the Bessel functions:
\begin{equation}
w=aJ_{\nu}\left(z\right)+bY_{\nu}\left(z\right)+H_{\nu}\left(z\right)\qquad\left(a,b\text{ constants}\right),\label{eq:w-Struve}
\end{equation}
where $z^{-\nu}H_{\nu}\left(z\right)$ is an integer function of $z$.
In particular and for concreteness, the power series expansion of Struve
function is given by:
\begin{equation}
H_{\nu}\left(z\right)=\left(\frac{1}{2}z\right)^{\nu+1}\sum_{k=0}^{\infty}\frac{\left(-1\right)^{k}\left(\frac{1}{2}z\right)^{2k}}{\Gamma\left(k+\frac{3}{2}\right)\Gamma\left(k+\nu+\frac{3}{2}\right)}.\label{eq:Struve-H}
\end{equation}

From (\ref{eq:EsHomPlusEsPart}), (\ref{eq:EsHom}) and (\ref{eq:EsPart}):
\begin{equation}
E_{r}\left(\zeta\right)=b_{3}I_{0}\left(\zeta\right)+b_{4}K_{0}\left(\zeta\right)-\frac{\pi}{2}\frac{\sigma}{\epsilon_{0}}\frac{R}{a}L_{0}\left(\zeta\right)\qquad\left(\zeta>\frac{R}{a}\right)\label{eq:Es(b3,b4)out}
\end{equation}
We are interested in the the asymptotic behavior of the above function
as $\zeta\rightarrow\infty$. Ref. \cite{Abramowitz} (Section 12.2.6)
informs us:
\begin{equation}
L_{\nu}\left(z\right)-I_{-\nu}\left(z\right)\sim\frac{1}{\pi}\sum_{k=0}^{\infty}\frac{\left(-1\right)^{k+1}\Gamma\left(k+\frac{1}{2}\right)}{\Gamma\left(\nu+\frac{1}{2}-k\right)\left(\frac{z}{2}\right)^{2k-\nu+1}}\qquad\left(z\gg1\right)\label{eq:AbramoEq12.2.6}
\end{equation}
Let $S_{\nu}\left(z\right)$ be defined as the right-hand side of
the above equation. Hence,
\begin{equation}
L_{\nu}\left(z\right)\sim I_{-\nu}\left(z\right)+S_{\nu}\left(z\right)\qquad\left(z\gg1\right).\label{eq:L(I,S)}
\end{equation}
Accordingly, $L_{0}\left(z\right)\sim I_{0}\left(z\right)+S_{0}\left(z\right)$.
This is enough to show that\ref{subsec:B-Field}
\begin{equation}
\lim_{z\rightarrow\infty}L_{0}\left(z\right)\rightarrow\infty\label{eq:L0_largez}
\end{equation}
since $\lim_{z\rightarrow\infty}I_{0}\left(z\right)\rightarrow\infty$---cf.
Eq. (\ref{eq:I0_K0_largez}). Notice that $S_{\nu}$ is finite in
the asymptotic limit $z\rightarrow\infty$.

Therefore, $E_{r}\left(\zeta\right)$ will diverge unless both the
first and third term in (\ref{eq:Es(b3,b4)out}) subtract to a finite
value. The combination of these terms is only possible through the
choice
\begin{equation}
b_{3}=\frac{\pi}{2}\frac{\sigma}{\epsilon_{0}}\frac{R}{a}.\label{eq:b3-out}
\end{equation}
In the face of this, Eq. (\ref{eq:Es(b3,b4)out}) reduces to 
\begin{equation}
E_{r}=b_{4}K_{0}\left(\zeta\right)-\frac{\pi}{2}\frac{\sigma}{\epsilon_{0}}\frac{R}{a}\left[L_{0}\left(\zeta\right)-I_{0}\left(\zeta\right)\right]\qquad\left(\zeta>\frac{R}{a}\right),\label{eq:eq:Es(b4)out}
\end{equation}
depending on the single integration constant $b_{4}$.

The complete solution for the electric field is assembled by considering
the inside solution (\ref{eq:Es(b1)in}) and the outside solution
(\ref{eq:eq:Es(b4)out}) : 
\begin{equation}
E_{r}\left(\zeta\right)=\begin{cases}
b_{1}I_{0}\left(\zeta\right)\qquad\left(\zeta<\frac{R}{a}\right) & \left(\text{inside the tube}\right)\\
b_{4}K_{0}\left(\zeta\right)-\frac{\pi}{2}\frac{\sigma}{\epsilon_{0}}\frac{R}{a}\left[L_{0}\left(\zeta\right)-I_{0}\left(\zeta\right)\right]\qquad\left(\zeta>\frac{R}{a}\right) & \left(\text{outside the tube}\right)
\end{cases}\label{eq:Es(b1,b4)}
\end{equation}
The integrations constants $b_{1}$ and $b_{4}$ should be determined
after suitable boundary conditions are applied.

The first boundary condition is given by demanding continuity of the
electric field on the tube's surface at $r=R$, i.e. $\zeta=R/a$:
$E_{r}^{\text{in}}\left(R/a\right)=E_{r}^{\text{out}}\left(R/a\right)$.
From (\ref{eq:Es(b1,b4)}), this maps to the equation:
\begin{equation}
b_{1}I_{0}\left(R/a\right)=b_{4}K_{0}\left(R/a\right)-\frac{\pi}{2}\frac{\sigma}{\epsilon_{0}}\frac{R}{a}\left[L_{0}\left(R/a\right)-I_{0}\left(R/a\right)\right]\qquad\left(\text{1st boundary condition}\right)\label{eq:E-1st-boundary-cond}
\end{equation}

The next boundary condition is to demand that continuity of the first-order
derivative of the electric field at the tube: $\frac{dE_{r}^{\text{in}}\left(R/a\right)}{d\zeta}=\frac{dE_{r}^{\text{out}}\left(R/a\right)}{d\zeta}$.
The derivatives of the Bessel functions are given by $\frac{dI_{0}}{d\zeta}=I_{1}\left(\zeta\right)$
and $\frac{dK_{0}}{d\zeta}=-K_{1}\left(\zeta\right)$; besides:

\begin{equation}
\frac{dL_{0}}{d\zeta}=L_{1}\left(\zeta\right)+\frac{2}{\pi}.\label{eq:dL0}
\end{equation}
This comes from a result involving the integral of $L_{1}\left(t\right)$;
for details we point the reader to Section 12.2.9 of \cite{Abramowitz}.
Also, the paragraph 12.2.1 of the same reference,
\begin{equation}
L_{\nu}\left(z\right)=-ie^{-i\nu\pi/2}H_{\nu}\left(z\right)=\left(\frac{1}{2}z\right)^{\nu+1}\sum_{k=0}^{\infty}\frac{\left(\frac{1}{2}z\right)^{2k}}{\Gamma\left(k+\frac{3}{2}\right)\Gamma\left(k+\nu+\frac{3}{2}\right)},\label{eq:ModifiedStruve-L}
\end{equation}
makes it clear that $L_{1}\left(0\right)=0$. Thus, using the expression
of the electric field (\ref{eq:Es(b1,b4)}), one finds:
\begin{equation}
b_{1}I_{1}\left(R/a\right)=-b_{4}K_{1}\left(R/a\right)-\frac{\pi}{2}\frac{\sigma}{\epsilon_{0}}\frac{R}{a}\left[L_{1}\left(R/a\right)-I_{1}\left(R/a\right)+\frac{2}{\pi}\right]\qquad\left(\text{2nd boundary condition}\right)\label{eq:E-2nd-boundary-cond}
\end{equation}

\begin{figure}
\begin{center}

\includegraphics[scale=0.5]{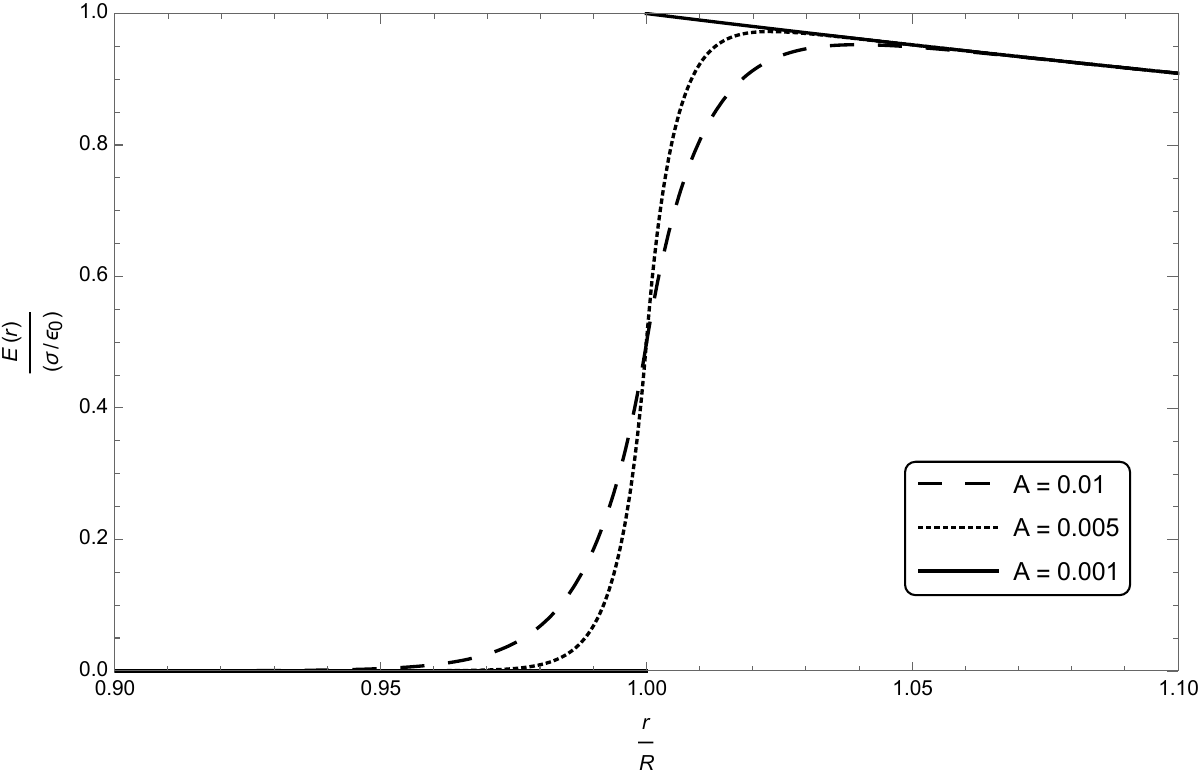}\end{center}

\caption{Behavior of the electric field produced by a infinite hole tube in
the context of Generalized Electrodynamics.}
\label{fig:E}
\end{figure}

The solution to the system formed by Eqs. (\ref{eq:E-1st-boundary-cond})
and (\ref{eq:E-2nd-boundary-cond}) provide the integration constants
$b_{1}$ and $b_{4}$. (While computing $b_{1}$ and $b_{4}$ we made
use of the Wronskian $W\left(R/a\right)=I_{0}\left(R/a\right)K_{1}\left(R/a\right)+I_{1}\left(R/a\right)K_{0}\left(R/a\right)=\frac{a}{R}$.)
Plugging the thus found $\left(b_{1},b_{2}\right)$ into Eq. (\ref{eq:Es(b1,b4)})
yields:
\begin{equation}
E_{r}\left(\zeta\right)=\begin{cases}
-\frac{\pi}{2}\frac{\sigma}{\epsilon_{0}}\left(\frac{R}{a}\right)^{2}\left\{ K_{1}\left(\frac{R}{a}\right)\left[L_{0}\left(\frac{R}{a}\right)-I_{0}\left(\frac{R}{a}\right)\right]+K_{0}\left(\frac{R}{a}\right)\left[L_{1}\left(\frac{R}{a}\right)-I_{1}\left(\frac{R}{a}\right)+\frac{2}{\pi}\right]\right\} I_{0}\left(\zeta\right) & \left(\zeta<\frac{R}{a}\right)\\
\\
-\frac{\pi}{2}\frac{\sigma}{\epsilon_{0}}\left(\frac{R}{a}\right)^{2}\left\{ I_{0}\left(\frac{R}{a}\right)\left[L_{1}\left(\frac{R}{a}\right)-I_{1}\left(\frac{R}{a}\right)+\frac{2}{\pi}\right]-I_{1}\left(\frac{R}{a}\right)\left[L_{0}\left(\frac{R}{a}\right)-I_{0}\left(\frac{R}{a}\right)\right]\right\} K_{0}\left(\zeta\right)+\\
-\frac{\pi}{2}\frac{\sigma}{\epsilon_{0}}\frac{R}{a}\left[L_{0}\left(\zeta\right)-I_{0}\left(\zeta\right)\right] & \left(\zeta>\frac{R}{a}\right)
\end{cases}\label{eq:Es(z)}
\end{equation}
This is the desired solution for the electric field produced by a
long tube of static charges in the context of GE. Fig. \ref{fig:E}
displays curves for $E_{r}\left(\zeta\right)$ for multiple values
of $A\equiv a/R$ as a function of the the distance from the tube's
axis in units of its radius, $r/R$.

Notice that the plots in Fig. \ref{fig:E} could have being built
directly from the expression (\ref{eq:Es(z)}) for $E_{r}\left(\zeta\right)$
and the power series expansion for the modified Struve function $L_{\nu}\left(z\right)$,
Eq. (\ref{eq:ModifiedStruve-L}).\footnote{The software Mathematica (used to construct the plots) has built-in
expressions for the Bessel functions $K_{\nu}$ and $I_{\nu}$ but
not for the modified Struve function $L_{\nu}$.} However, that was not the way we plotted the curves in Fig. \ref{fig:E}.
Instead, we considered Eqs. (\ref{eq:AbramoEq12.2.6}) and (\ref{eq:L(I,S)})
to express $L_{\nu}\left(z\right)$ in terms of $S_{\nu}\left(z\right)$.
This is a fare strategy because $R\gg a$ (for physical consistency),
so that the condition for large $z$ is satisfied and 
\begin{equation}
L_{\nu}\left(z\right)-I_{-\nu}\left(z\right)\sim\frac{1}{\pi}\sum_{k=0}^{\infty}\frac{\left(-1\right)^{k+1}\Gamma\left(k+\frac{1}{2}\right)}{\Gamma\left(\nu+\frac{1}{2}-k\right)\left(\frac{z}{2}\right)^{2k-\nu+1}}\equiv S_{\nu}\left(z\right)\qquad\left|z\right|\gg1\label{eq:S_nu}
\end{equation}
is actually valid. Then, $L_{\nu}\left(z\right)\sim S_{\nu}\left(z\right)+I_{-\nu}\left(z\right)$
and we can take: $L_{0}\left(z\right)=S_{0}\left(z\right)+I_{0}\left(z\right)$
and $L_{1}\left(z\right)=S_{1}\left(z\right)+I_{1}\left(z\right)$.
Here, the result $I_{-\nu}\left(z\right)=I_{\nu}\left(z\right)$ for
$\left(\nu\text{ integer}\right)$ was utilized. In this way, the
coefficients of the functions $I_{0}\left(\zeta\right)$ and $K_{0}\left(\zeta\right)$
appearing in Eq. (\ref{eq:Es(z)}) simply, leading to:
\begin{equation}
E_{r}\left(r\right)=\begin{cases}
-\frac{\pi}{2}\frac{\sigma}{\epsilon_{0}}\left(\frac{R}{a}\right)^{2}\left[K_{1}\left(\frac{R}{a}\right)S_{0}\left(\frac{R}{a}\right)+K_{0}\left(\frac{R}{a}\right)S_{1}\left(\frac{R}{a}\right)+\frac{2}{\pi}K_{0}\left(\frac{R}{a}\right)\right]I_{0}\left(\frac{r}{a}\right) & \left(r\leqslant R\right)\\
-\frac{\pi}{2}\frac{\sigma}{\epsilon_{0}}\left(\frac{R}{a}\right)^{2}\left[I_{0}\left(\frac{R}{a}\right)S_{1}\left(\frac{R}{a}\right)-I_{1}\left(\frac{R}{a}\right)S_{0}\left(\frac{R}{a}\right)+\frac{2}{\pi}I_{0}\left(\frac{R}{a}\right)\right]K_{0}\left(\frac{r}{a}\right)-\frac{\pi}{2}\frac{\sigma}{\epsilon_{0}}\frac{R}{a}S_{0}\left(\frac{r}{a}\right) & \left(r>R\right)
\end{cases},\label{eq:Es(z,S_nu)}
\end{equation}
where $\zeta=\frac{r}{a}$ and $R\gg a$ is assumed. It was Eq. (\ref{eq:Es(z,S_nu)})
that produced the curves in Fig. \ref{fig:E}.

\bigskip{}

That Eq. (\ref{eq:Es(z,S_nu)}) is the correct solution can be verified
in a number of ways, including: (i) by checking continuity of $E_{r}$
at $r=R$; and (ii) by computing the asymptotic limits to Maxwell
ordinary predictions. We leave the proof of (i) for the interested
reader. Checking (ii) is much more interesting from the physical point
of view. In fact, we use the asymptotic forms from paragraphs 9.7.1
and 9.7.2 of Ref. \cite{Abramowitz},
\begin{equation}
I_{\nu}\left(z\right)\simeq\frac{e^{z}}{\sqrt{2\pi z}}\qquad\text{and}\qquad K_{\nu}\left(z\right)\sim\sqrt{\frac{\pi}{2z}}e^{-z}\qquad\left(z\gg1\right),\label{eq:I_nu_K_nu_largez}
\end{equation}
and
\begin{equation}
S_{0}\left(z\right)\approx-\frac{1}{z}\frac{2}{\pi}\left(1+\frac{1}{z^{2}}\right)\qquad\text{and}\qquad S_{1}\left(z\right)\approx-\frac{2}{\pi}\left(1-\frac{1}{z^{2}}\right)\qquad\left(z\gg1\right),\label{eq:S_0_S_1_largez-1}
\end{equation}
The latter result stems from keeping the leading terms in the sum
defining $S_{\nu}\left(z\right)$---see Eq. (\ref{eq:S_nu}). The
approximations (\ref{eq:I_nu_K_nu_largez}) and (\ref{eq:S_0_S_1_largez-1})
are useful within the square brackets in both lines of Eq. (\ref{eq:Es(z,S_nu)})
because the argument of $I_{0}$, $I_{1}$, $K_{0}$, $K_{1}$, $S_{0}$
and $S_{1}$ therein is $\frac{R}{a}\gg1$. Under (\ref{eq:I_nu_K_nu_largez})
and (\ref{eq:S_0_S_1_largez-1}), Eq. (\ref{eq:Es(z,S_nu)}) reduces
to:
\begin{equation}
E_{r}\left(\zeta\right)=\begin{cases}
\frac{\sigma}{\epsilon_{0}}\sqrt{\frac{\pi R}{2a}}e^{-R/a}I_{0}\left(\zeta\right)\qquad\left(\zeta<\frac{R}{a}\right) & \left(\text{inside the tube}\right)\\
-\frac{\sigma}{\epsilon_{0}}\sqrt{\frac{R}{2\pi a}}e^{R/a}K_{0}\left(\zeta\right)+\frac{\sigma}{\epsilon_{0}}\frac{R}{a}\frac{1}{\zeta}\qquad\left(\zeta>\frac{R}{a}\right) & \left(\text{outside the tube}\right)
\end{cases}.\label{eq:Es(z)_largez}
\end{equation}

The Maxwell limit can be obtained for the interior solution for $\zeta\approx0$,
while in the exterior it is necessary to take $\zeta\rightarrow\infty$.
According to paragraph 9.6.7 of Ref. \cite{Abramowitz}: $I_{\nu}\left(z\right)\sim\left(\frac{z}{2}\right)^{\nu}\frac{1}{\Gamma\left(\nu+1\right)}$
for $\left|z\right|\ll1$ as long as $\nu\neq-1,-2,\dots$. Therefore,
\begin{equation}
E_{r}^{\text{in}}\left(r\right)\sim\frac{\sigma}{\epsilon_{0}}\sqrt{\frac{\pi R}{2a}}e^{-R/a},\qquad\left|\zeta\right|\ll1\qquad\left(\text{inside Maxwell limit}\right)\label{eq:Es_in_M}
\end{equation}
which is damped down to zero since $R\gg a$ for the reasonable physical
situation where Podolsky is only a small correction to Maxwell. The
exterior solution far from the tube will be given by the second line
of (\ref{eq:Es(z)_largez}) and (\ref{eq:I_nu_K_nu_largez}):
\begin{equation}
E_{r}^{\text{out}}\left(r\right)\sim-\frac{\sigma}{2\epsilon_{0}}\sqrt{\frac{R}{r}}e^{-\left(r-R\right)/a}+\frac{\sigma}{\epsilon_{0}}\frac{R}{r}\qquad\left|\zeta\right|\gg1\qquad\left(\text{outside Maxwell limit}\right).\label{eq:Es_out_M}
\end{equation}
Herein, the first term is the Podolsky correction, which decays from
the surface on, and the last term is exactly the Maxwell solution.\footnote{By the way, the forthcoming Eq. (\ref{eq:phi(E)}), leads to the conclusion
that $\varphi^{\text{out}}=-\int E_{r}^{\text{out}}dr=-\frac{\sigma}{\epsilon_{0}}R\ln\frac{r}{R}$
after the decaying of the first term in Eq. (\ref{eq:Es_out_M}).
This is entirely consistent with the electric potential outside of
a charged tube in Maxwell electrodynamics \cite{GriffithsEletro}. }

\subsection{Electric potential for a long charged tube in GE \label{subsec:V-potential}}

As we have said before (see the beginning of Section \ref{subsec:A-Potential}),
GE is U(1)-gauge invariant. Therefore, the electric potential $\varphi$
can be calculated via
\begin{equation}
\mathbf{E}=-\nabla\varphi+\frac{\partial\mathbf{A}}{\partial t}\label{eq:E(phi,A)-coda}
\end{equation}
just like in Maxwell electrodynamics. Indeed, Eq. (\ref{eq:E(phi,A)-coda})
follows from GE field equations (\ref{eq:FieldEqsGE}); in particular,
it stems from the second and third equations in (\ref{eq:FieldEqsGE}).

Let us calculate the electric potential $\varphi$ from the expression
of electric field in the previous section. Recall that the vector
potential $\mathbf{A}$ is null for the configuration at hand: a long
tube of static charges. Then, Eq. (\ref{eq:E(phi,A)-coda}) is simply:
\begin{equation}
\mathbf{E}=-\nabla\varphi\Rightarrow E_{r}=-\frac{\partial\varphi}{\partial r}.\label{eq:E(phi)}
\end{equation}
The last step comes from the fact that $\mathbf{E}=\left(E_{r},0,0\right)$,
cf. (\ref{eq:SymmetryOnE}). Integrating (\ref{eq:E(phi)}) gives:
\begin{equation}
\varphi=-\int E_{r}dr=-a\int E_{r}\left(\zeta\right)d\zeta,\label{eq:phi(E)}
\end{equation}
where 
\begin{equation}
E_{r}\left(\zeta\right)=\begin{cases}
b_{1}I_{0}\left(\zeta\right)\qquad\left(\zeta<\frac{R}{a}\right) & \left(\text{inside the tube}\right)\\
b_{4}K_{0}\left(\zeta\right)-\frac{\pi}{2}\frac{\sigma}{\epsilon_{0}}\frac{R}{a}\left[L_{0}\left(\zeta\right)-I_{0}\left(\zeta\right)\right]\qquad\left(\zeta\geqslant\frac{R}{a}\right) & \left(\text{outside the tube}\right)
\end{cases}\label{eq:Es(b1,b4)-coda}
\end{equation}
from Eq. (\ref{eq:Es(b1,b4)}). Recall that $\zeta=\frac{r}{a}$ as
usual.

The integrals of the modified Bessel and Struve functions are given
in terms of a hypergeometric functions:\footnote{The results in Eqs. (\ref{eq:int_I0})-(\ref{eq:int_L0}) where given
by software Mathematica. See also Ref. \cite{Abramowitz}.}
\begin{equation}
\int I_{0}\left(\zeta\right)d\zeta=\zeta\,{}_{1}F_{2}\left(\frac{1}{2};1,\frac{3}{2};\frac{\zeta^{2}}{4}\right);\label{eq:int_I0}
\end{equation}
\begin{equation}
\int K_{0}\left(\zeta\right)d\zeta=\frac{\pi}{2}\zeta\,\left[K_{0}\left(\zeta\right)L_{-1}\left(\zeta\right)+K_{1}\left(\zeta\right)L_{0}\left(\zeta\right)\right];\label{eq:int_K0}
\end{equation}
\begin{equation}
\int L_{0}\left(\zeta\right)d\zeta=\frac{\zeta^{2}}{\pi}\,{}_{2}F_{3}\left(1,1;\frac{3}{2},\frac{3}{2},2;\frac{\zeta^{2}}{4}\right).\label{eq:int_L0}
\end{equation}

The potential inside the tube is computed from Eqs. (\ref{eq:phi(E)}),
the first line in (\ref{eq:Es(b1,b4)-coda}), and (\ref{eq:int_I0}).
Analogously, the potential outside the tube comes from Eqs. (\ref{eq:phi(E)}),
(\ref{eq:Es(b1,b4)-coda}) line two, (\ref{eq:int_K0}), and (\ref{eq:int_L0}).The
complete expression for the electric potential will be:
\begin{equation}
\varphi\left(r\right)=\begin{cases}
-b_{1}r\,{}_{1}F_{2}\left(\frac{1}{2};1,\frac{3}{2};\frac{1}{4}\frac{r^{2}}{a^{2}}\right)+\varphi_{0}^{\text{in}} & \left(r<R\right)\\
-b_{4}\frac{\pi}{2}r\,\left[K_{0}\left(\frac{r}{a}\right)L_{-1}\left(\frac{r}{a}\right)+K_{1}\left(\frac{r}{a}\right)L_{0}\left(\frac{r}{a}\right)\right]\\
\quad-\frac{\pi}{2}\frac{\sigma}{\epsilon_{0}}\frac{R}{a}\left[\frac{1}{\pi}\frac{r^{2}}{a^{2}}\,{}_{2}F_{3}\left(1,1;\frac{3}{2},\frac{3}{2},2;\frac{1}{4}\frac{r^{2}}{a^{2}}\right)-\frac{r}{a}\,{}_{1}F_{2}\left(\frac{1}{2};1,\frac{3}{2};\frac{1}{4}\frac{r^{2}}{a^{2}}\right)\right]+\varphi_{0}^{\text{out}} & \left(r\geqslant R\right)
\end{cases},\label{eq:phi(r)}
\end{equation}
where $b_{1}$ and $b_{4}$ are given by 
\begin{equation}
b_{1}=-\frac{\pi}{2}\frac{\sigma}{\epsilon_{0}}\left(\frac{R}{a}\right)^{2}\left\{ K_{1}\left(\frac{R}{a}\right)\left[L_{0}\left(\frac{R}{a}\right)-I_{0}\left(\frac{R}{a}\right)\right]+K_{0}\left(\frac{R}{a}\right)\left[L_{1}\left(\frac{R}{a}\right)-I_{1}\left(\frac{R}{a}\right)+\frac{2}{\pi}\right]\right\} \label{eq:b1}
\end{equation}
and
\begin{equation}
b_{4}=-\frac{\pi}{2}\frac{\sigma}{\epsilon_{0}}\left(\frac{R}{a}\right)^{2}\left\{ -I_{1}\left(\frac{R}{a}\right)\left[L_{0}\left(\frac{R}{a}\right)-I_{0}\left(\frac{R}{a}\right)\right]+I_{0}\left(\frac{R}{a}\right)\left[L_{1}\left(\frac{R}{a}\right)-I_{1}\left(\frac{R}{a}\right)+\frac{2}{\pi}\right]\right\} ,\label{eq:b4}
\end{equation}
 cf. explained in section \ref{subsec:E-field} below Eq. (\ref{eq:E-2nd-boundary-cond}).

\subsection{Electric Aharonov-Bohm phase shift in GE\label{sec:Electric-AB-phase-shift}}

Consider again the setup in Fig. \ref{fig:ABsetup_Electric}. Let
the upper-left (bottom-right) cylinder be called tube 1 (tube 2) kept
at an electric potential $\varphi_{1}$ $\left(\varphi_{2}\right)$.
The beam of charged particles is split into two branches before passing
through the tubes; branch 1 (branch 2) is approximately linear when
passing through tube 1 (tube 2) and is regarded parallel to its axis,
although it might not coincide with it. In fact, for generality we
shall take branch 1 (branch 2) at a distance $r_{1}$ (distance $r_{2}$)
of the axis of tube 1 (tube 2). This detail in the preparation of
the beam splitting will be crucial for the detection of the Aharonov-Bohm
effect in GE as opposed to the same effect in regular Maxwell electrodynamics. 

\begin{figure}
\begin{center}

\includegraphics[scale=0.5]{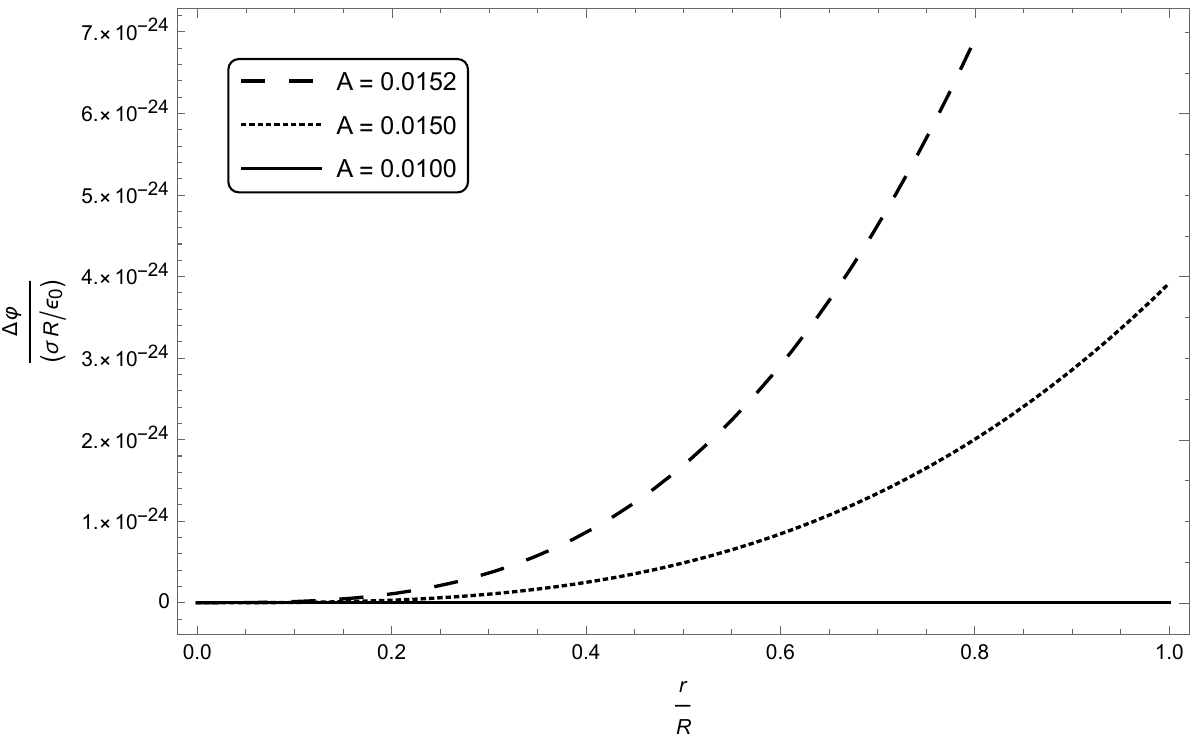}

\end{center}

\caption{Behavior of the electric potential difference produced by two infinite
hole tubes in the context of Generalized Electrodynamics.}
\label{fig:Delta-phi}
\end{figure}

The electric potential difference $\Delta\varphi$ is built from Eq.
(\ref{eq:phi(r)}). Actually, only the first line of this equation
(potential internal to the tube) will matter for computing the electric
AB effect in GE since the beam traverse the inwards of the tubes.
Then,
\begin{align}
\Delta\varphi\left(\zeta_{1},\zeta_{2},R_{1},R_{2},\sigma_{1},\sigma_{2}\right) & =\varphi_{2}\left(\zeta_{2}\right)-\varphi_{1}\left(\zeta_{1}\right)\nonumber \\
 & =-b_{1}\left(R_{2},\sigma_{2}\right)\left(a\zeta_{2}\right)\,{}_{1}F_{2}\left(\frac{1}{2};1,\frac{3}{2};\frac{\zeta_{2}^{2}}{4}\right)+\varphi_{0,2}^{\text{in}}\nonumber \\
 & \qquad+b_{1}\left(R_{1},\sigma_{1}\right)\left(a\zeta_{1}\right)\,{}_{1}F_{2}\left(\frac{1}{2};1,\frac{3}{2};\frac{\zeta_{1}^{2}}{4}\right)-\varphi_{0,1}^{\text{in}}.\label{eq:Delta_phi}
\end{align}
Notice that $\Delta\varphi$ is a function of the distance $\zeta=r/a$
separating each beam branch of its corresponding tube's axis; it also
depends on the radii $R_{1}$ and $R_{2}$ of tube 1 and tube 2, and
of the surface charge density $\sigma$ for each tube. In principle,
$R_{1}\neq R_{2}$, $\sigma_{1}\neq\sigma_{2}$, and so on; however,
the Podolsky coupling $a$ remains the same in both tubes for obvious
reasons. Eq. (\ref{eq:Delta_phi}) has the form 
\begin{equation}
\Delta\varphi=\Delta\varphi^{\left(M\right)}+\Delta\varphi^{\left(P\right)}.\label{eq:Delta_phi_M_P}
\end{equation}
The regular Maxwellian contribution is
\begin{equation}
\Delta\varphi^{\left(M\right)}=\varphi_{0,2}^{\text{in}}-\varphi_{0,1}^{\text{in}},\label{eq:Delta_phi_M}
\end{equation}
where $\varphi_{0,i}^{\text{in}}$ is the constant Maxwell potential
in the interior of the $i$-th tube. Podolsky contribution to the
potential difference is dubbed $\Delta\varphi^{\left(P\right)}$;
it is given in terms of the hypergeometric function $_{1}F_{2}$ and
of Eq. (\ref{eq:b1}) for $b_{1}$. In order to simplify the latter
expression, let both tubes be identical, i.e., with same radius $R_{1}=R_{2}=R$
and same surface charge density $\sigma_{1}=\sigma_{2}=\sigma$; even
so, we keep the beam branches at different positions with respect
to the axis of their respective tubes $\left(r_{1}\neq r_{2}\right)$.
Then, Podolsky correction to the electric potential difference reads:
\begin{align}
\Delta\varphi^{\left(P\right)}\left(r_{1},r_{2}\right) & =\frac{\pi}{2}\left(\frac{\sigma}{\epsilon_{0}}R\right)\left(\frac{R}{a}\right)^{2}\nonumber \\
 & \quad\times\left\{ K_{1}\left(\frac{R}{a}\right)\left[L_{0}\left(\frac{R}{a}\right)-I_{0}\left(\frac{R}{a}\right)\right]+K_{0}\left(\frac{R}{a}\right)\left[L_{1}\left(\frac{R}{a}\right)-I_{1}\left(\frac{R}{a}\right)+\frac{2}{\pi}\right]\right\} \nonumber \\
 & \quad\times\left[\frac{r_{2}}{R}\,{}_{1}F_{2}\left(\frac{1}{2};1,\frac{3}{2};\frac{1}{4}\frac{R^{2}}{a^{2}}\frac{r_{2}^{2}}{R^{2}}\right)-\frac{r_{1}}{R}\,{}_{1}F_{2}\left(\frac{1}{2};1,\frac{3}{2};\frac{1}{4}\frac{R^{2}}{a^{2}}\frac{r_{1}^{2}}{R^{2}}\right)\right].\label{eq:Delta_phi_P(z)}
\end{align}
Eq. (\ref{eq:Deltag_AB}) gives the phase shift 
\begin{equation}
\Delta g=-\frac{q}{c\hbar}\oint\varphi dx^{0}=-\frac{qt}{\hbar}\Delta\varphi\label{eq:Deltag_AB_Electric}
\end{equation}
where $\Delta\varphi$ is given by (\ref{eq:Delta_phi_M_P}), (\ref{eq:Delta_phi_M})
and (\ref{eq:Delta_phi_P(z)}). Therefore, the plots in Fig. \ref{fig:Delta-phi}
inform directly the behavior of the complete phase shift $\Delta g$,
given a time interval $t$ (and the charge $q$ of the particles in
the beam).

\bigskip{}

As before, let $A=a/R$ where $A\ll1$ for assessing small Podosky
corrections to Maxwell electrodynamics. The expression (\ref{eq:Delta_phi_P(z)})
presents numerical problems when one tries to plot the function $\Delta\varphi^{\left(P\right)}\left(r_{1},r_{2}\right)$
in the range $r/R\in\left\{ 0,1\right\} $ for $A\ll1$. In order
to bypass this issue, we perform a power expansion of the electric
field close to the inner radius of the tube:
\begin{equation}
E_{r}\left(\zeta\right)=b_{1}I_{0}\left(\zeta\right)\approx b_{1}\left[I_{0}\left(1\right)+\left.\frac{dI_{0}}{d\zeta}\right\vert _{\zeta=1}\left(\zeta-1\right)+\frac{1}{2}\left.\frac{d^{2}I_{0}}{d\zeta^{2}}\right\vert _{\zeta=1}\left(\zeta-1\right)^{2}+O\left(\zeta^{3}\right)\right]\label{eq:Es-in-expansion}
\end{equation}
with
\begin{equation}
\left.\frac{dI_{0}}{d\zeta}\right\vert _{\zeta=1}=I_{1}\left(1\right)\qquad\text{and}\qquad\left.\frac{d^{2}I_{0}}{d\zeta^{2}}\right\vert _{\zeta=1}=I_{0}\left(1\right)-I_{1}\left(1\right).\label{eq:I0_derivatives}
\end{equation}
Hence, from (\ref{eq:phi(E)}), (\ref{eq:Es-in-expansion}) and (\ref{eq:I0_derivatives}):
\begin{equation}
\varphi^{\text{in}}\left(r\right)\simeq-Rb_{1}\left\{ I_{0}\left(1\right)\zeta+I_{1}\left(1\right)\frac{\left(\zeta-1\right)^{2}}{2}+\frac{1}{2}\left[I_{0}\left(1\right)-I_{1}\left(1\right)\right]\frac{\left(\zeta-1\right)}{3}^{3}\right\} .\label{eq:phi-in-expansion}
\end{equation}

Now, the electric potential difference (\ref{eq:Delta_phi_P(z)})
between the two identical tubes becomes:
\begin{align}
\frac{\Delta\varphi^{\left(P\right)}\left(S_{1},S_{2}\right)}{\left(\frac{\sigma}{\epsilon_{0}}R\right)} & =\frac{\pi}{2}\left(\frac{1}{A}\right)^{2}\nonumber \\
 & \quad\times\left\{ K_{1}\left(\frac{1}{A}\right)\left[L_{0}\left(\frac{1}{A}\right)-I_{0}\left(\frac{1}{A}\right)\right]+K_{0}\left(\frac{1}{A}\right)\left[L_{1}\left(\frac{1}{A}\right)-I_{1}\left(\frac{1}{A}\right)+\frac{2}{\pi}\right]\right\} \nonumber \\
 & \quad\times\left\{ I_{0}\left(1\right)\left(\frac{S_{2}}{A}-\frac{S_{1}}{A}\right)+\frac{1}{2}I_{1}\left(1\right)\left[\left(\frac{S_{2}}{A}-1\right)^{2}-\left(\frac{S_{1}}{A}-1\right)^{2}\right]+\right.\nonumber \\
 & \qquad\left.+\frac{1}{6}\left[I_{0}\left(1\right)-I_{1}\left(1\right)\right]\left[\left(\frac{S_{2}}{A}-1\right)^{3}-\left(\frac{S_{1}}{A}-1\right)^{3}\right]\right\} \label{eq:Delta_phi_P-expansion}
\end{align}
Using the asymptotic limit given in Eq. (\ref{eq:Es(z)_largez})
for $R\gg a$, we have
\begin{align}
\frac{\Delta\varphi^{\left(P\right)}\left(S_{1},S_{2}\right)}{\left(\frac{\sigma}{\epsilon_{0}}R\right)} & =\sqrt{\frac{\pi}{2A}}e^{-1/A}\left\{ I_{0}\left(1\right)\frac{S_{2}}{A}+I_{1}\left(1\right)\frac{\left(\frac{S_{2}}{A}-1\right)^{2}}{2}+\frac{1}{2}\left[I_{0}\left(1\right)-I_{1}\left(1\right)\right]\frac{\left(\frac{S_{2}}{A}-1\right)}{3}^{3}\right.\nonumber \\
 & \left.-I_{0}\left(1\right)\frac{S_{1}}{A}-I_{1}\left(1\right)\frac{\left(\frac{S_{1}}{A}-1\right)^{2}}{2}-\frac{1}{2}\left[I_{0}\left(1\right)-I_{1}\left(1\right)\right]\frac{\left(\frac{S_{1}}{A}-1\right)}{3}^{3}\right\} \label{eq:Delta_phi_P-largez}
\end{align}
Fig. \ref{fig:Delta-phi} shows the plots for $\Delta\varphi^{\left(P\right)}$
in Eq. (\ref{eq:Delta_phi_P-largez}) with $S_{1}=0$ (the beam branch
1 travels along the axis of tube number 1) and $S_{2}=S=r/R$. As
expected, the smaller the $A$, the closer to Maxwell predictions. 

\bigskip{}

In the limit $R\gg a$, we could even use Eq. (\ref{eq:Es_in_M})
for $E_{r}^{\text{in}}$:
\begin{equation}
\varphi^{\text{in}}\left(s\right)=-\int E_{r}^{\text{in}}\left(\zeta\right)Rd\zeta=-\frac{\sigma}{\epsilon_{0}}\sqrt{\frac{\pi R}{2a}}re^{-R/a}+\varphi_{0}^{\text{in}}\label{eq:phi_in_Rgga}
\end{equation}
where $\varphi_{0}^{\text{in}}=\varphi^{\text{in}}\left(0\right)$.
In this approximate case,
\begin{align}
\Delta\varphi & =\varphi_{2}^{\text{in}}\left(r_{2}\right)-\varphi_{1}^{\text{in}}\left(r_{1}\right)=\nonumber \\
 & =\left(\frac{\sigma}{\epsilon_{0}}R\right)\sqrt{\frac{\pi}{2}\frac{R}{a}}\left(\frac{r_{1}}{R}-\frac{r_{2}}{R}\right)e^{-R/a}+\Delta\varphi^{\left(M\right)}.\label{eq:Delta_phi_Rgga}
\end{align}
Herein we considered equal-sized tubes $\left(R_{1}=R_{2}=R\right)$
equipped with the same charge distribution $\left(\sigma_{1}=\sigma_{2}=\sigma\right)$;
the regular Maxwell contribution is denoted by $\Delta\varphi^{\left(M\right)}\equiv\varphi_{0,2}^{\text{in}}-\varphi_{0,1}^{\text{in}}$.

It is important to note that, in the case of equally charged identical
tubes with beams passing through the same distance off the center
$\left(r_{1}=r_{2}\right)$, the additional Podolsky term vanishes.
Also, if the beam branches travel exactly along the center of each
tube, there will be no contribution from Generalized Electrodynamics
to the electric Aharonov-Bohm effect. 

\section{Final Comments \label{sec:Final-Comments}}

This paper dealt with the analysis of the magnetic Aharonov-Bohm effect
and the electric Aharonov-Bohm effect in the contest of Bopp-Podolsky's
Generalized Electrodynamics. The phase shifts due to the interaction
of the beam of particles with the magnetic vector potential and the
electric potential were computed and the diferences with their Maxwellian
counterparts stressed. 

In order to do so, we followed a similar route in both the magnetic
case and the electric case. Specifically, the magnetic field $\mathbf{B}$
(electric field $\mathbf{E}$) was computed by solving Ampére-Maxwell-Podolsky
law (Gauss-Podolsky law) for a stationary current (static charge sistribution)
in a long solenoid (a long cylinder)---see Eq. (\ref{eq:FieldEqsGE}).
Faraday law and the equation for the absence of magnetic monopoles
are the same in GE as in Maxwell electrodynamics; this is the case
because GE is also U(1) gauge invariant \cite{Cuzinatto2007}. Therefore,
the definition of magnetic field $\mathbf{B}$ (of the electric field
$\mathbf{E}$) in terms of the the vector potential $\mathbf{A}$
(of the electric potential $\varphi$ and of $\mathbf{A}$) in the
contexts of GE and of Maxwell electrodynamics is the same, i.e. $\mathbf{B}=\nabla\times\mathbf{A}$
$\left(\mathbf{E}=-\nabla\varphi+\frac{\partial\mathbf{A}}{\partial t}\right)$.
The limits of the fields and potentials from GE to Maxwell theory
were verified each step of the way.

Once $\mathbf{A}$ and $\varphi$ for GE were determined by direct
integration of the field equations, we moved on to compute the phase
shifts $\Delta g$ according to the path integral technique summarized
by Felsager \cite{Felsager}. It was proven that Generalized Electrodynamics
produces modification in $\Delta g$ with respect to the Maxwellian
prediction $\Delta g^{\left(M\right)}$. In fact, in the limit as
$R\gg a$, the magnetic case is described by Eqs. (\ref{eq:Deltag(Deltag(M))})
and (\ref{eq:deltag(P)}),
\begin{equation}
\frac{\Delta g}{\Delta g^{\left(M\right)}}=1-\left(\frac{a}{R}\right)\left(\frac{r}{R}\right)^{1/2}e^{-\left(r-R\right)/a}\qquad\left(\text{magnetic AB in GE}\right),\label{eq:Deltag-magnetic}
\end{equation}
while in the electric case is given by Eq. (\ref{eq:phi_in_Rgga}),
\begin{equation}
\Delta g=\Delta\varphi^{\left(M\right)}+\left(\frac{\sigma}{\epsilon_{0}}R\right)\sqrt{\frac{\pi}{2}\frac{R}{a}}\left(\frac{r_{1}}{R}-\frac{r_{2}}{R}\right)e^{-R/a}\qquad\left(\text{electric AB in GE}\right).\label{eq:Deltag-electric}
\end{equation}
The new contributions from GE are present in the last terms of Eqs.
(\ref{eq:Deltag-magnetic}) and (\ref{eq:Deltag-electric}). We see
that these terms are exponentially suppressed since $R\gg a$, the
radius of the macroscopic cylinders are much larger than GE parameter
$a$. The latter should be submicroscopic in order to Podolsky's corrections
to Maxwell theory be small, as required by extensive tests of the
standard electrodynamics \cite{Luo2005,Goldhaber2010}. In fact, Podolsky
mass ${\color{blue}m_{\gamma}=\hbar\left(ca\right)^{-1}}$ for the
massive mode of the photon was constrained to $\apprge370\text{ GeV}$
\cite{Bufalo2014}.\textcolor{blue}{{} According to this energy value,
Podolsky's length parameter $a=\hbar c\left(m_{\gamma}c^{2}\right)^{-1}$
should be $\approx0.53\times10^{-18}$ m or less. This value agrees
with the bound given by Carley et al. in Ref. \cite{Carley2019},
which argues that precision measurements of the Lymann-$\alpha$ line
in the Schrödinger spectrum of a hydrogen atom suggest that Bopp-Landé-Thomas-Podolsky
length $a$ must be smaller than $\approx10^{-18}$ m.}\footnote{\textcolor{blue}{See also the estimate in Ref. \cite{Cuzinatto2011}
for comparison.}}\textcolor{blue}{{} This is at least three orders of magnitude smaller
than the empirical proton radius (about 0.85 fm) \cite{Brandt2022}.}

In the face of what was said, the perspective of constraining $a$
\textcolor{blue}{(an atto-metre length scale parameter)} via Aharonov-Bohm
effect measurements \textcolor{blue}{is not achievable with the current technology}. This is
true also in the best-case scenario, where $a$ could be assessed
in a the beam flyby version of the magnetic AB effect, cf. discussed
at the end of section \ref{sec:Magnetic-AB-phase-shift}. Even so,
the present work is satisfying from the theoretical point of view;
it analyses a key effect in the interface of electrodynamics, gauge
theory and quantum mechanics. At the same time, it adds an application
to Generalized Electrodynamics, an alternative theory that has been
intensely explored in the recent years \cite{Frenkel1996,Frenkel1999,Cuzinatto2011,Bonin2010,Bufalo2014,Carley2019,Carley2023,Bufalo2011,Bufalo2012,Bufalo2013,Brandt2016,Lazar2019,Cuzinatto2018,Frizo2023,Cuzinatto2017}.

\section*{Acknowledgements}

CAMM and RRC thank FAPEMIG-Brazil (Grants APQ-00544-23 and APQ-0528-23)
for partial financial support. BMP and JCSE acknowledge CAPES-Brazil
for financial support. RRC is also grateful to CNPq-Brazil (Grants
309984/2020-3 and 309063/2023-0) for partial financial support. \textcolor{blue}{The
authors are indebted to the reviewer, whose comments helped improve
the physics discussion in this paper.}

\end{document}